\newcommand{\st}{(\xvec{s},t)}
\newcommand{\sth}{(\xvec{s},t,h)}
\newcommand{\h}{(h)}
\newcommand{\spline}[1]{\xvec{\phi}_{#1}^T\h}
\newcommand{\bs}{bike-sharing}
\newcommand{\bss}{\bs\ system}
\newcommand{\bsss}{\bss s} 
\newcommand{\ci}{confidence interval}
\newcommand{\cip}[1]{$#1\%$-confidence interval}
\newcommand{\bhat}[1]{$\hat{\xvec{\beta}}_{\text{#1}}\h$}
\newcommand{\bootpar}{\gamma} 
\newcommand{\clock}{~o'clock}

\documentclass[12pt, a4paper]{article}
\usepackage[utf8]{inputenc}
\usepackage{dcolumn,lscape}
\usepackage{amsmath,longtable,multicol,dcolumn,tabularx,graphicx,amssymb}
\usepackage{exscale,amsthm,multirow,rotating,subcaption}
\usepackage{natbib}
\usepackage{bbm}
\usepackage{hyperref}
\usepackage{tikz,dsfont,float}
\newcommand{\xvec}{\boldsymbol}
\newcommand{\xmat}{\mathbf}

\usetikzlibrary{patterns}

\begin{document}

\def\spacingset#1{\renewcommand{\baselinestretch}%
{#1}\small\normalsize} \spacingset{1}


\title{\bf A Spatiotemporal Functional Model for Bike-Sharing Systems - An Example based on the City of Helsinki}
\author{Andreas Piter\\
{\small Leibniz University Hannover, Germany}\\
  Philipp Otto\footnote{Corresponding author: philipp.otto@uni-goettingen.de}\\
    {\small University of Göttingen, Germany}\\
    Hamza Alkhatib\\
    {\small Leibniz University Hannover, Germany}}
  \maketitle

\begin{abstract}
Understanding the usage patterns for bike-sharing systems is essential in terms of supporting and enhancing operational planning for such schemes. Studies have demonstrated how factors such as weather conditions influence the number of bikes that should be available at bike-sharing stations at certain times during the day. However, the influences of these factors usually vary over the course of a day, and if there is good temporal resolution, there could also be significant effects only for some hours/minutes (rush hours, the hours when shops are open, and so forth). Thus, in this paper, an analysis of Helsinki's bike-sharing data from 2017 is conducted that considers full temporal and spatial resolutions. Moreover, the data are available at a very high frequency. Hence, the station hire data is analysed in a spatiotemporal functional setting, where the number of bikes at a station is defined as a continuous function of the time of day. For this completely novel approach, we apply a functional spatiotemporal hierarchical model to investigate the effect of environmental factors and the magnitude of the spatial and temporal dependence. Challenges in computational complexity are faced using a bootstrapping approach. The results show the necessity of splitting the bike-sharing stations into two clusters based on the similarity of their spatiotemporal functional observations in order to model the station hire data of Helsinki's bike-sharing system effectively. The estimated functional influences of the proposed factors are different for the two clusters. Moreover, the estimated parameters reveal high random effects in the data that are not explained by the mean of the process. In this random-effects model, the temporal autoregressive parameter dominates the spatial dependence.
\end{abstract}

\noindent%
\vfill

\spacingset{1.45} 

\section{Introduction}\label{sec:introduction}

Bicycle-sharing systems have become popular in all the cities in which they have been implemented. Among these cities is the capital of Finland, Helsinki, where a station-based system has provided a flexible transport option since 2016 \citep{HelBikes,HSLBikesPlans}. That is, a bike can be taken from a \bs\ station and returned to any other station. The extension of the existing system to neighbouring cities and the rebalancing of the \bs\ stations are challenging tasks for the operator and city planners (cf. \citealt{schuijbroek2017inventory}). Support can be provided via empirical research examining the demand at all stations over time, and determining the factors that influence the usage of a \bss \ is one of the main research interests. Furthermore, the environmental impact of implementing bike-sharing schemes is an important question in current research (e.g. \citealt{maranzano2020analysis,zhang2018environmental}).

In studies concerning \bsss\ in other cities, a variety of influential factors have been discussed and analysed using different statistical models \citep{wang2020spatiotemporal,el2017effects,yang2016mobility}. The significance and magnitudes of such factors have been analysed by \cite{eren2020review} in a review. However, studies have analysed and predicted the \bs\ usage at different points in time (e.g., \citealt{yang2020exploring}). Furthermore, the amount of data is often reduced through the aggregation of the observations in time spans consisting of a set number of hours (e.g., \citealt{el2017effects}) or a day (e.g., \citealt{buck2012bike}).
So far, no study has analysed how factors’ influences vary by time of day. Additionally, no study has utilized an entire dataset in its analyses, i.e., all the studies to date have aggregated the data or reduced the dimensionality.

This gap in knowledge is addressed in this paper by means of a comprehensive analysis of the freely available station-hire data for the \bss\ in Helsinki from 2017. The station-hire data is meant to represent spatiotemporal functional observations. Thus, we apply a complex spatiotemporal functional hierarchical model implemented in the software package D-STEM (cf. \citealt{finazzi2020dstem2,fasso2018statistical,finazzi2014dstem}).
This model can be used to predict and map the spatiotemporal process and its uncertainty over a geographical region across time. Applications of such dynamic corregionalisation models are used in \cite{fasso2016european,fasso2013varying,finazzi2013model,taghavi2019concurrent} to assess the air quality in Europe and model the concentrations of several airborne pollutants in a multivariate setting or for land use regression in Teheran, Iran. In contrast to these models, which handle the purely temporal dynamics separately from the purely spatial correlation component, the approach presented in \cite{calculli2015maximum} combines the spatial and temporal dependencies in an autoregressive spatial component. It is known as hidden dynamic geostatistical model (HDGM). The parameters are estimated using the maximum likelihood approach and an EM algorithm (cf. \citealt{finazzi2014dstem}). Alternatively, Bayesian approaches can also be used (cf. \citealt{Rue09}). These approaches are mostly based on computationally efficient integrated nested Laplace approximations (INLA). However, we focus on the EM estimation for  functional hierarchical models implemented in D-STEM. Although this technique was originally developed to handle spatiotemporal functional data from environmental sciences, such as atmospheric radiosonde profiles, its potential for modelling the number of allocated bikes at the \bsss\ in Helsinki is demonstrated in this paper.

Because of the large amount of data from 140 stations (measured in five-minute intervals), we propose to combine this estimation with a bootstrap approach. Thus, we will be able to estimate the standard errors of the estimated functional model parameters in a very efficient way, allowing a rich interaction model with spatiotemporal interactions to be estimated from the full data. Moreover, all results are validated in a cross-validation study.

The remainder of the paper is structured as follows. First, we introduce the spatiotemporal functional model from a theoretical perspective and explain the applied bootstrapping principle. The concept of functional data and the construction of a continuous function from discrete observations are described. In Section \ref{sec:literature}, an overview of the field of \bs\ analysis is provided via a thorough literature review. These theoretical sections are followed by the empirical analysis of the data from Helsinki. Initially, we discuss several descriptive statistics and figures in detail to provide a comprehensive understanding of the data, which is highly complex (i.e., spatial, temporal domain; daily, weekly periodicity; high frequency; and so forth). Eventually, the estimated functional parameters are shown and the results are discussed in Section \ref{sec:results}. In this section, we also explain how the specific model (hyper-)parameters are chosen. Section~\ref{sec:conclusion} concludes the paper.

\section{Modelling spatiotemporal dependence in functional data}\label{sec:model}

Functional data analysis deals with a functional random variable~$Y$ that is continuously defined, as in, e.g ., \cite{ferraty2006nonparametric}. Observations~$y$ of a functional random variable are either measured on a regular grid or at random discrete points, thus leading to a set of $q$~discrete measurements $y_{i,1},\ldots,y_{i,q}$ of the functional data $i\in {1,\ldots,m}$. It is worth noting that $q$ can be different for different functional data~$y_i$. However, for functional data analysis, a continuous function is needed to evaluate the function $y(h)$ for any argument~$h$. Hence, the functional form is reconstructed, for instance, using the basis function expansion (cf. \citealt{ndongo2017spatio,wang2016functional}). That is, the reconstruction is accomplished with a set of $K$ known basis functions~$\phi_k$ with respective coefficients~$c_k$, $y(h)$ that can be expressed as
\begin{eqnarray}
y(h) = \sum_k^K \phi_k(h) c_k = \xvec{\phi}^T(h) \xvec{c}.
\end{eqnarray}
Typical choices for the basis functions are the Fourier series for periodic data or the B-splines for non-periodic data  \citep{ramsay1997functional}. For this analysis, we focus on the B-spline approach, where the number of free parameters $p$ is given by the order of the piecewise polynomials and the number of interior knots.
The compact support of the B-spline basis functions has the advantage that the computational complexity increases only linearly with~$K$. Furthermore, B-spline basis functions are flexible in the sense that the location of the break points can be chosen in order to approximate the function better in segments where it changes more frequently.

Let $Y\sth$ be a functional space-time random variable, and let $y\sth$ be the observed functions from $h \in [a,b] \subseteq \mathds{R}$ to $\mathds{R}$ at time~$t$ and spatial location~$\xvec{s}\in D$, with $D\subset S^2$ and $S^2$ being the unit sphere in $\mathds{R}^3$. Time is assumed to be discrete, with $t\in\{1,\dots,T\}$. Furthermore, the actual observations of $Y\sth$ are made at $q$ discrete points along the dimension of the function $y\sth$. Hence, the observation at $\st$ is the $q$-dimensional vector $\xvec{y}\st=(y_1\st,\dots,y_q\st)^T$. The spatiotemporal functional variable $Y\sth$ is assumed to be first-order stationary (i.e., the mean of the spatiotemporal process does not depend on the location).

A hierarchical model is used to model the mean spatiotemporal functional process and its variation by splitting up the total uncertainty into separate components. The first level is given by
\begin{eqnarray}
&& y\sth = \mu\sth+\omega\sth+\varepsilon\sth \, , \label{eq:hierarchical_model}
\end{eqnarray}
where the $\mu\sth$ are fixed effects, and the $\omega\sth$ are spatially and temporally correlated random effects. The model errors $\varepsilon\st$ are assumed to be from a Gaussian white noise process with a constant variance that is allowed to vary over the functional domain. More precisely,
\begin{eqnarray}
\varepsilon\sth\sim N(0,\tilde{\sigma}^2\h) \, ,\label{eq:measurement_error}
\end{eqnarray}
with
\begin{eqnarray}
&& \tilde{\sigma}^2\h = \spline{\varepsilon} \xvec{\sigma}_\varepsilon^2 .
\end{eqnarray}
It is important to note that the spline basis functions could be chosen differently for each term. Hence, the basis functions $\phi^T_a$ and their dimension~$p_a$ are denoted by the subscript corresponding to the terms $a \in \{\mu,\omega,\varepsilon\}$ of the hierarchical model given in~\eqref{eq:hierarchical_model}. 

The fixed effect model
\begin{eqnarray}
&& \mu\sth = \displaystyle\sum_{i=1}^{d} x_{\mu,i}\sth \spline{\mu} \xvec{\beta}_i \label{eq:fixedeffect} 
\end{eqnarray}
consists of $d$ space-time varying functional covariates $x_{\mu,i}\sth$, where the unknown coefficients $\xvec{\beta}_i$ must be estimated. It is worth noting that these covariates could also be constant across space or time and/or in the functional dimension. Further, the random-effects model is given by
\begin{eqnarray}
&& \omega\sth = \spline{\omega} \xvec{z}\st . \label{eq:randomeffect} 
\end{eqnarray}
It covers both spatial and temporal dependencies by modelling the respective variation using a basis function expansion. Specifically, the spatiotemporal latent component $\xvec{z}\st$ has the Markovian dynamics
\begin{eqnarray}
&& \xvec{z}\st = \xmat{G}\xvec{z}(\xvec{s},t-1) + \xvec{\eta}\st  \, , \label{eq:markovianDynamics}
\end{eqnarray}
with $\xvec{\eta}(t)$ being the spatially dependent Gaussian process
\begin{eqnarray}
&& \xvec{\eta}(t) \sim N(0, \xmat{V} \otimes  \rho(||\xvec{s}-\xvec{s}^\prime||,\theta,\nu)). \label{eq:etaGaussianProcess}
\end{eqnarray}
Here, $\otimes$ stands for the Kronecker product. While the cross-covariance matrix $\xmat{V}$ describes the correlation between all components, the spatial covariance function $\rho(||\xvec{s}-\xvec{s}^\prime||, \theta, \nu)$ captures the correlation across space. For instance, this function could be a Matérn covariance function, which is an isotropic covariance function depending on the distance $||\xvec{s}-\xvec{s}^\prime||$ between spatial locations only, for the matrix $\xmat{V} = diag(\sigma_{\eta_1}^2, \dots, \sigma_{\eta_{p_\omega}}^2)$. The second restriction is made mainly to reduce the computational effort. To estimate the parameters, we follow the maximum likelihood approach using an EM algorithm implemented in D-STEM (see \citealt{finazzi2020dstem2}), which uses a functional hierarchical model called the f-HDGM model. For more details on the closed form and the numerical computations of the parameters in the EM algorithm, we refer the reader to \cite{finazzi2020dstem2,calculli2015maximum,fasso2011maximum,fasso2009algorithm}.

However, for this approach, the computational costs increase drastically with the number of spatial locations~$n$ and the number of splines~$p_\omega$ chosen for the basis function expansion. This is due to the inversion of the $np_\omega\times np_\omega$ covariance matrix of the spatially dependent Gaussian process given in \eqref{eq:etaGaussianProcess}. In our study of Helsinki's bike-sharing system, the number of bikes at 140 stations was observed every five minutes for 176 days. Thus, we propose to use a bootstrapping procedure to estimate the standard errors of the model parameters more efficiently (see, e.g., \citealt{efron1986bootstrap}).

Suppose that~$Y\sth \sim F_\bootpar$ with an unknown probability distribution~$F_\bootpar$ for which there are observations $\{ y\sth \}$, as above. In general, there is a parameter of interest called $\bootpar (F_\bootpar)$ (model parameters in our case) that cannot be computed directly, as the probability distribution $F_\bootpar$ is unknown. However, $\bootpar (F_\bootpar)$ is approximated by the bootstrap estimate $\hat{\bootpar} = \bootpar (\hat{F}_\bootpar)$, with $\hat{F}_\bootpar$ being the empirical distribution. This  estimate is computed in three steps. First, $B$ independent bootstrap samples $\xvec{b}_1,\ldots,\xvec{b}_B \in \{\xvec{s} \in D: y\sth\}$ are generated, each consisting of $m$~locations that are randomly drawn with replacement from all locations where the process is being observed. Then the parameter of interest is estimated separately for each of the $B$ bootstrap samples.
The estimator of the $i$-th bootstrap sample $\{ y\sth : \xvec{s} \in \xvec{b}_i \}$ is denoted by $\hat{\bootpar}_i$. Lastly, $\hat{\bootpar}$ is given by
\begin{eqnarray}
\hat{\bootpar} = \frac{1}{B} \sum_{i=1}^B \hat{\bootpar}_i.
\end{eqnarray}
Further, the bootstrap standard error
\begin{eqnarray}
\hat{\sigma} = \sqrt{\frac{1}{B-1}\sum_{i=1}^B(\hat{\bootpar}_i-\hat{\bootpar})^2}
\end{eqnarray}
is the non-parametric maximum likelihood estimate of the true standard error. This bootstrap approach is non-parametric in the sense that no assumptions are made about the unknown distribution~$F_\bootpar$.
Finally, the confidence interval of the bootstrap estimate $\hat{\bootpar}$ can be computed using the percentile method. Here, the $(1-\alpha)$ confidence interval is approximated by the respective empirical quantiles of the bootstrap distribution of $\hat{\bootpar}$.

\section{Bike-sharing systems}\label{sec:literature}

Many of the larger cities across the world have expanded their public transport systems by introducing bike-share schemes, which provide an alternative and sustainable transport mode. Starting in 2000, the number of bike-share systems worldwide has increased rapidly, and many studies have been conducted to improve these services and understand their usage patterns \citep{gervini2019exploring,fishman2016bikeshare}.

The systems differ in the way they handle bike usage  \citep{eren2020review}. On the one hand, there are station-based systems, where users retrieve a bike from a particular station, take a ride and return it to any other station. Here, a positive effect is that the station locations are fixed and thus users know where to search for bikes. However, a bike station may already be full when a user wants to return a bike, as there are only a limited number of docks at each station. Then the bike can only be returned to another station.
On the other hand, dockless sharing systems offer more flexibility, as users can return bikes anywhere and they are not bound to stations. However, the disadvantage is that users who want to pick up a bike need to be lucky to find a bike close by. Bike-sharing systems have become popular for various reasons \citep{o2014mining}. City administrations aim at increasing the number of cyclists and reducing the car traffic in the cities \citep{fishman2016bikeshare}. Shared bikes can be used to overcome distances between public transport options, such as the metro or train, to reach specific destinations, such as the workplace or recreational areas. Hence, bike sharing improves the public transport network and helps users cover gaps in that network or the last miles \citep{willberg2019bike}.

Research in this area proceeds in various directions but mainly aims to understand users’ behaviour and the different facets of bike-sharing demand. The knowledge gained from the investigations helps improve bike-sharing systems and support operators in operational planning. Understanding usage patterns of, and dependencies between, stations may help when introducing similar systems to other cities \citep{tran2015modeling}. Moreover, \cite{martinez2012optimisation,garcia2012optimizing} address finding appropriate station locations and determining bike fleet size. Data from user registrations or from user surveys provides the users’ perspective and gives insights into the socio-demographic factors influencing the usage of bike-sharing systems \citep{willberg2019bike}.

A different focus is set by data-driven demand analysis. On the one hand, there is station hire data, which give either the number of bikes or the number of check-outs and check-ins at each station at a certain point in time. On the other hand, trip-based data gives information about the origin, destination and duration of each bicycle trip. Thus, there are several studies that investigate spatial and temporal factors influencing the demand on a station level or trip basis that try to predict future usage \citep{yang2016mobility,rixey2013station,li2015traffic}.
However, station-based bike-sharing systems suffer from an unbalanced spatial distribution of bikes at the stations due to different levels of demand across space and time. Hence, this optimization problem must find the most effective rebalancing strategy for the bikes in the network \citep{shi2019study}.

A recently published literature review by \cite{eren2020review} on the factors influencing bike-sharing demand focuses on six categories. The categories used are weather conditions, built environment, public transport and temporal factors that are used in many studies on station hire data. One of the main findings is that precipitation affects bike-sharing demand the most among the meteorological covariates. Its negative correlation with demand was found in almost all examined studies. Furthermore, increases in the humidity and wind speed decrease the demand, whereas air temperatures between $0-30^\circ$C lead to more bicycle trips. The strongest positive correlation was found for temperatures between $20-30^\circ$C, but the demand is less for temperatures under $0^\circ$C and over $30^\circ$C. Infrastructure and land use are widely investigated factors in the built environment category. Bicycle lanes and the proximity of bike-sharing stations to them are found to have high positive impacts  on a station's demand.
Furthermore, changes in the elevation across the area of a bike-sharing service are correlated with its demand. From trip-based data, it can be seen that users tend to use shared bikes to go downhill more than uphill. Moreover, considerable differences in demand are found for bike-sharing stations in commercial and residential areas, and a station's proximity to infrastructure, such as museums, shopping centres, schools, universities and restaurants, is investigated by many studies.
Also, public transport options seem to be related to the bike-sharing demand. The more train, tram, bus and metro stations near a station, the higher its demand. Moreover, many studies have shown that the bike-sharing demand varies along the temporal dimension, with the most apparent differences occurring between weekdays and weekends due to different user travel motivations. The usage of bike-sharing for commuting to work becomes visible via the peak usage during morning and afternoon rush hours on weekdays. On the other hand, trips during the weekend are more often for recreational purposes.

Most of the studies use linear models (e.g., generalised linear models \citep{ludovic2017cycling}, hierarchical linear mixed effect models \citep{el2017effects} or negative binomial models \citep{gebhart2014impact,nair2013large}) without explicitly addressing spatial dependence. \cite{yang2020exploring,ji2018exploring,wang2020spatiotemporal} model the demand using ordinary least squares regression, which is inconsistent in the presence of spatial dependence, but they subsequently check residuals for spatial autocorrelation using Moran's I \citep{lee2017extending}. In other studies, spatial dependence is mostly addressed via cluster analyses (e.g., \citealt{lathia2012measuring,zhou2015understanding,froehlich2009sensing,raninen2018spatiotemporal,li2015traffic,vogel2011understanding}). In contrast, temporal dependence has been studied more accurately. For instance, \cite{shi2018exploring} studied metro riderships explicitly addressing its temporal dimension, while \cite{el2017effects} consider first-order temporal autoregressive model. In some studies, temporal dependence has been ruled out for the dependent/independent variables through aggregation over time, e.g., average number of trips per month \citep{rixey2013station}, per day \citep{buck2012bike} or during the peak hours in the morning or afternoon \citep{tran2015modeling,nair2013large,wang2020spatiotemporal}.

The city of Helsinki introduced the station-based  public bike-share scheme in 2016 with 50 stations and further expanded it in 2017 with 100 additional stations (see \citealt{HelBikes} and \citealt{jappinen2013modelling}). As a consequence of its high usage and popularity, the system was extended to the neighbouring cities Espoo (2018) \citep{HSLBikesPlans} and Vantaa (2019) \citep{HSLBikesVantaa}. Hence, the bike-sharing system covers wide areas of the larger Helsinki region and has become a dense network of bike-sharing stations, making this system an alternative transport mode. Helsinki's bike-sharing scheme has been addressed before in a few studies (see \citealt{tarnanen2017modelling,ludovic2017cycling,raninen2018spatiotemporal}).

\section{Empirical results}\label{sec:results}

In the following sections, bike-sharing usage in Helsinki is analysed in the spatiotemporal functional framework as described above. For this purpose, we initially take a closer look at the station hire data, thus highlighting the complexity and variety of the data. This investigation includes a spectral time series analysis showing the usefulness of a spatiotemporal functional dimension. In the ensuing sections, the results of the empirical model are discussed in more detail to show which behaviour patterns can be revealed by the statistical model.

\subsection{Descriptive statistics}

\begin{figure}
\centering
\includegraphics[width=0.33\textwidth]{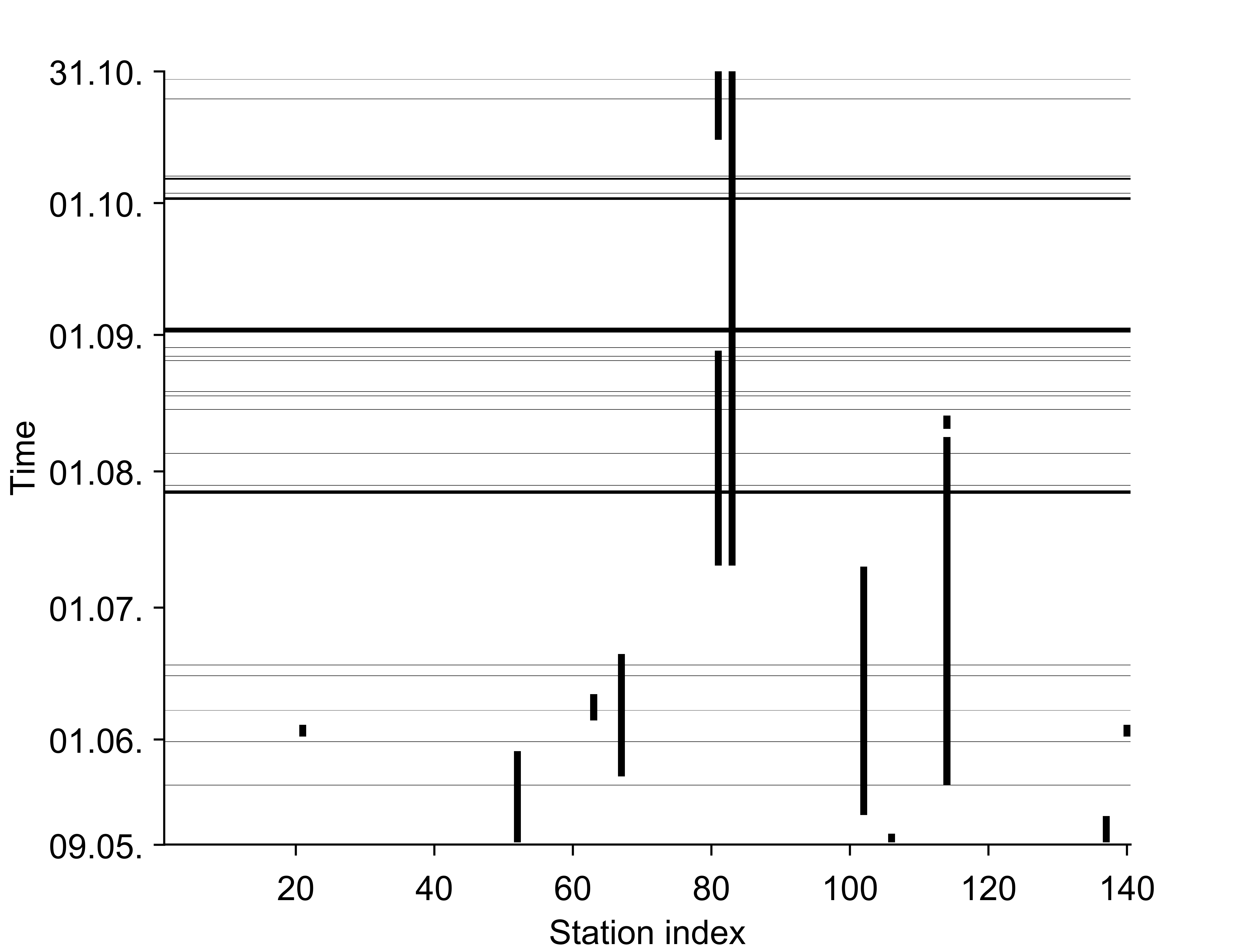}
\caption{Missing data in the station hire dataset from 2017 in Helsinki. Black lines show the data gaps over time (horizontal) and for certain \bs\ stations (vertical). The width of the horizontal lines corresponds to the length of the period that the data is missing.}
\label{fig:DataLack}
\end{figure}

The company \cite{HRT2020} has provided an API to enable individuals and organisations to develop their own applications and investigate the data related to transport in Helsinki and neighbouring municipalities  \cite[cf.][]{Kainu}. We selected 176 days in 2017,  starting with the $9^\text{th}$ May,  which was the first day with full records, and ending on the $31^\text{st}$ October, which appears to have been the end of the biking season that year. However, the data is incomplete, as shown in Figure~\ref{fig:DataLack}, where black entries depict missing values. Due to the hierarchical model structure and the estimation being based on an EM algorithm, data imputation is not necessary, as the model is capable of imputing these values (cf. \citealt{fasso2013varying,finazzi2014dstem}).

\begin{figure}
\centering
\begin{subfigure}{0.49\textwidth}
\centering
\includegraphics[width=1\textwidth]{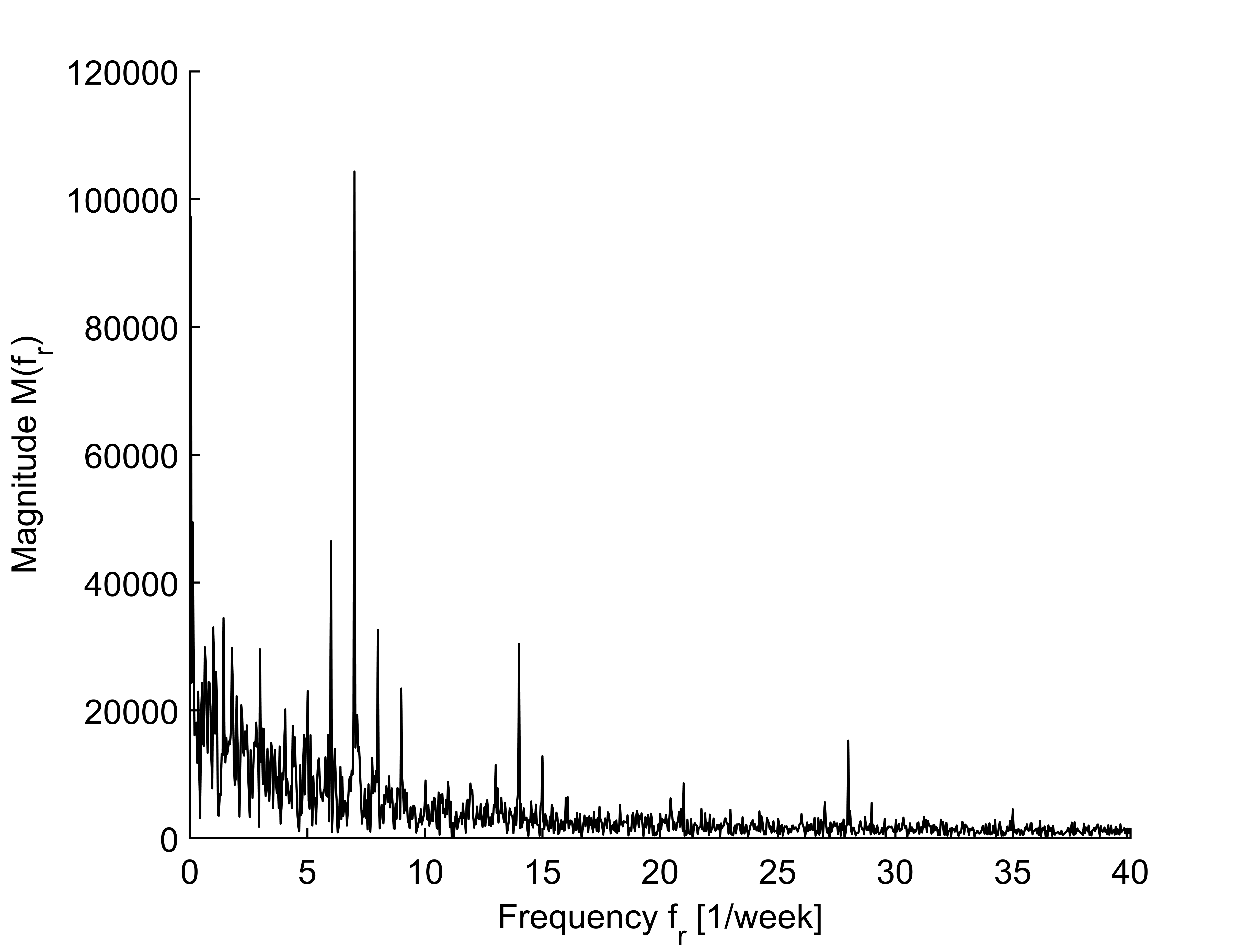}
\caption{Periodogram of \bs\ station \textit{Arabiankatu}, which serves as reference for the glyph-map.}
\label{fig:frequency_legend}
\end{subfigure}
\begin{subfigure}{0.49\textwidth}
\centering
\includegraphics[width=1\textwidth]{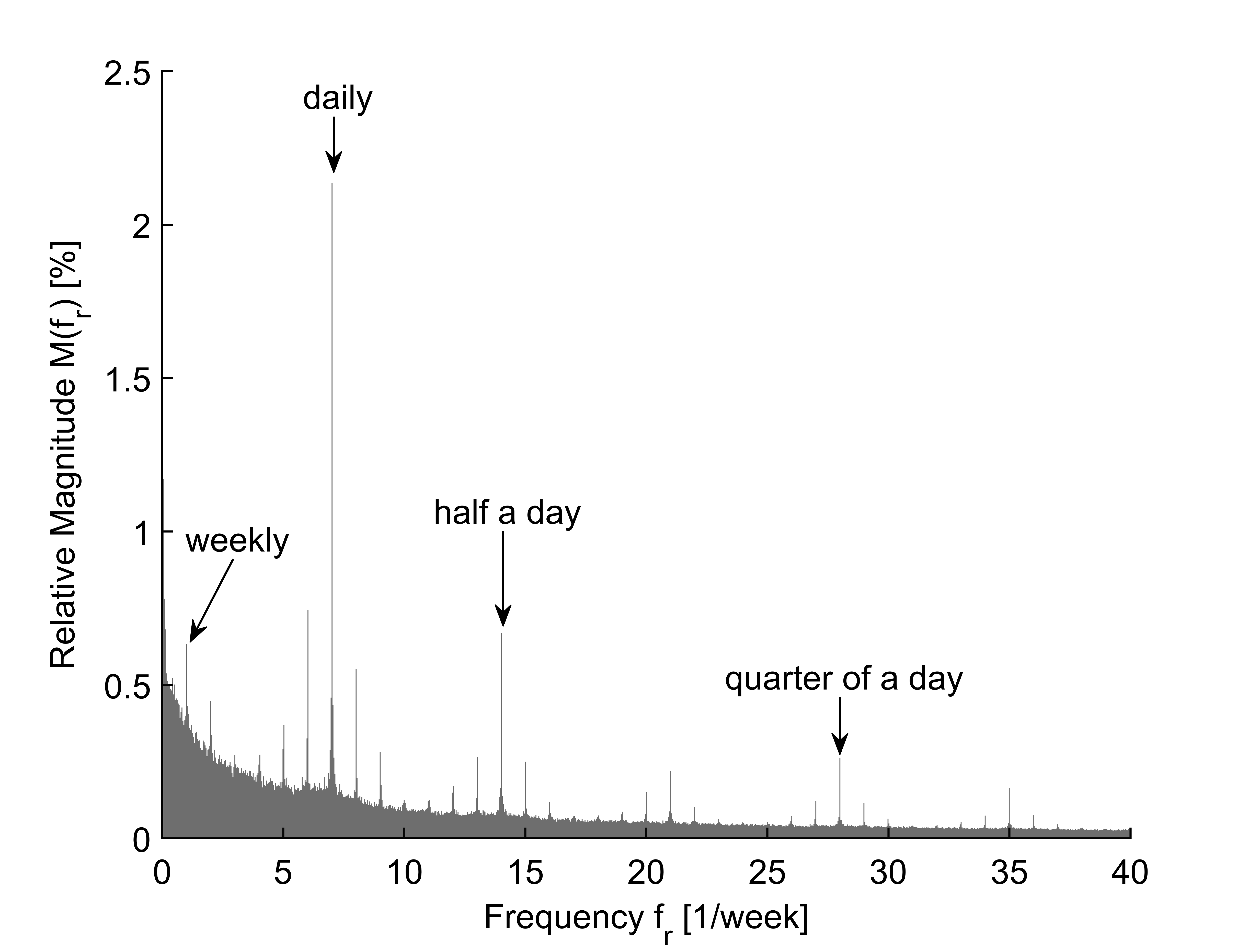}
\caption{Histogram of relative magnitudes of each frequency computed from the data for all \bs\ stations.}
\label{fig:frequency_histogram}
\end{subfigure}
\begin{subfigure}{0.7\textwidth}
\centering
\includegraphics[width=0.8\textwidth]{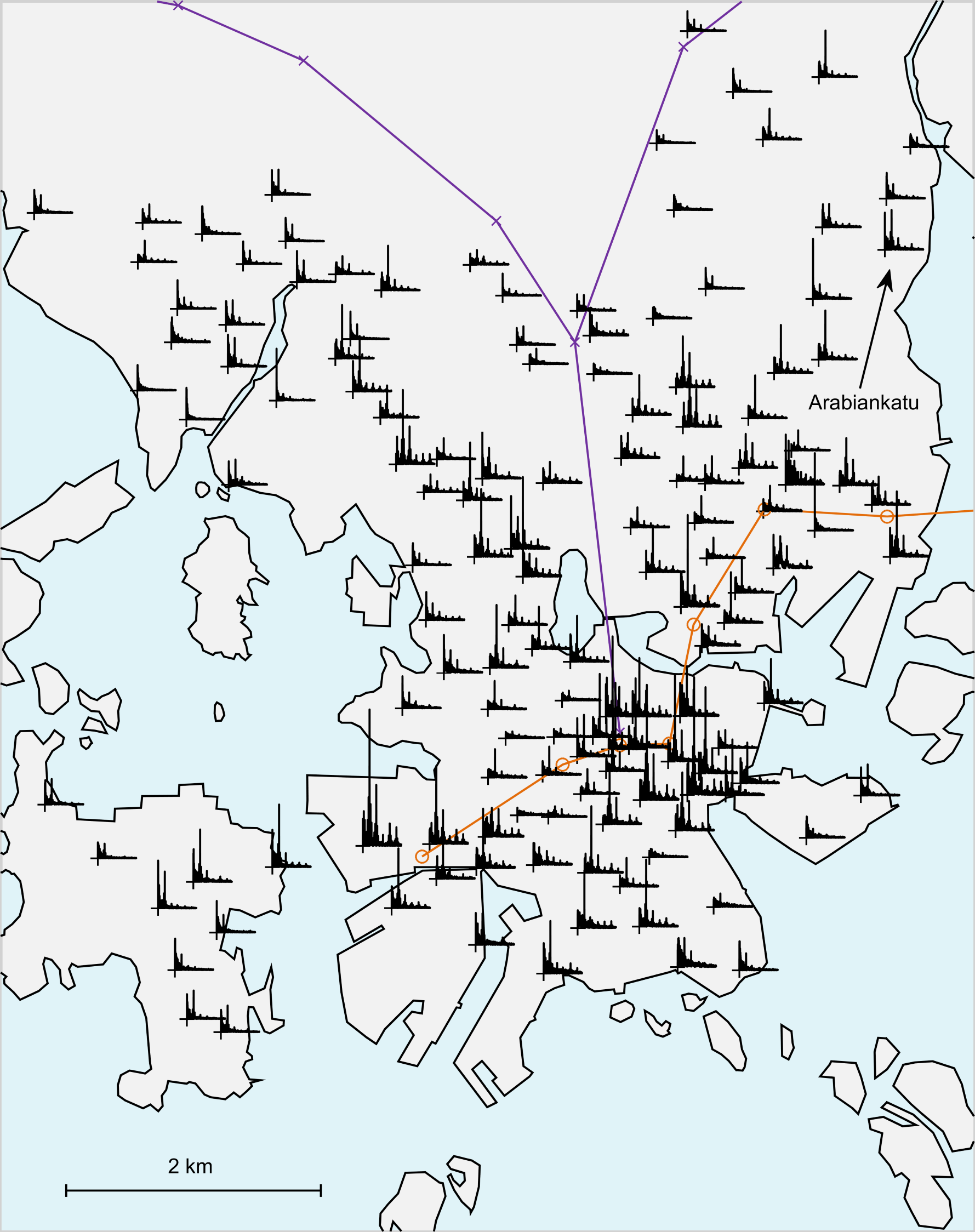}
\caption{Glyph-map.}
\label{fig:frequency_map}
\end{subfigure}
\caption{Glyph-map of the periodograms of  all the \bs\ stations in Helsinki, with the periodogram for \textit{Arabiankatu} used as a reference and the histogram showing relative magnitudes.}
\label{fig:glyphMap_frequency}
\end{figure}

The station hire data contains information on the observed number of bikes recorded every five minutes at 140 bike-share stations in Helsinki. Thus, there are over 7 million bicycle counts in total. In a preliminary analysis, the time series for the stations were analysed in the frequency domain (cf. \citealt{cooley1965algorithm,brockwell2016introduction}). Periodograms were computed for all the station separately and are depicted in Figure~\ref{fig:frequency_map} via a glyph-map \citep{wickham2012glyph,eden2010two} imposed on a reference plot of the stations \textit{Arabiankatu} from Figure~\ref{fig:frequency_legend}.  The periodograms are shown as small glyphs at the locations of the respective bike-sharing stations. Thus, Figure~\ref{fig:frequency_map} additionally illustrates the spatial distribution of the stations. Generally, the urban area of Helsinki is covered widely by bike-share stations, but they are denser and evenly distributed in the city centre. The coverage becomes sparse and irregular in the north of Helsinki where there are more residential areas present.

The glyph-map reveals differences in the periodograms that appear to be spatially correlated. In the city centre close to the main station, the magnitudes are higher than in most parts in the north of Helsinki. Moreover, the bike-sharing stations in the city centre have periodograms with several dominant peaks. However, most of the periodograms for the stations have one outstanding peak in common. This dominant peak corresponds to the daily frequency. This finding is further highlighted in the histogram for the relative magnitudes in Figure~\ref{fig:frequency_histogram}. The histogram shows an aggregation of all the periodograms from the glyph-map. Here, the magnitude~$M(f_r)\,[\%]$ of the $r$-th frequency~$f_r$ relative to the total signal of all stations is computed with
\begin{eqnarray}
M(f_r)\,[\%] = \frac{\sum_{i=1}^n M_i(f_r)}{\sum_{i=1}^n \sum_{j=1}^{N_f} M_i(f_j)}\cdot 100 \, ,
\end{eqnarray}
where ~$n$ is the number of stations, and ~$N_f$ is the number of frequencies in the periodogram.
According to the histogram in Figure~\ref{fig:frequency_histogram}, the daily cycle is most prominent, representing over 2\,$\%$ of the total signal from all stations.

In general, temporal data can be subdivided into linear time, e.g., a sequence of subsequent days, and cyclic time, e.g., daily, weekly or yearly periodicities \citep{andrienko2010space}. The station hire data contains both temporal types of data.
First of all, there are two predominant periodicities that can be understood as cyclic time. The most prominent cycle is the length of one day. Its prominence could be due to daily activity and sleep periods. Moreover, there is a cycle with a length of one week that is connected to the transition between workdays (Monday through Friday) and the weekend (Saturday and Sunday).
Nevertheless, these repetitive structures appear for a sequence of days, which represents linear time. Both types of time have to be considered in the analysis of the \bs\ data in order to cover all spatial and temporal dependencies within the data.
Therefore, this study makes the novel proposal that the cyclic time of one day should be treated as a functional observation (cf. \citealt{ferraty2006nonparametric,ramsay2007applied}). To be precise, the number of bikes is one continuous function of time across a day, i.e., a function that maps $[0,24]$ to $\mathds{N}_0$. Thus, the daily cycle is incorporated into the functional observations for every day at every station.

\begin{figure}
\begin{subfigure}{0.49\textwidth}
  \centering
  \includegraphics[width=1\linewidth,trim= 100 0 100 0]{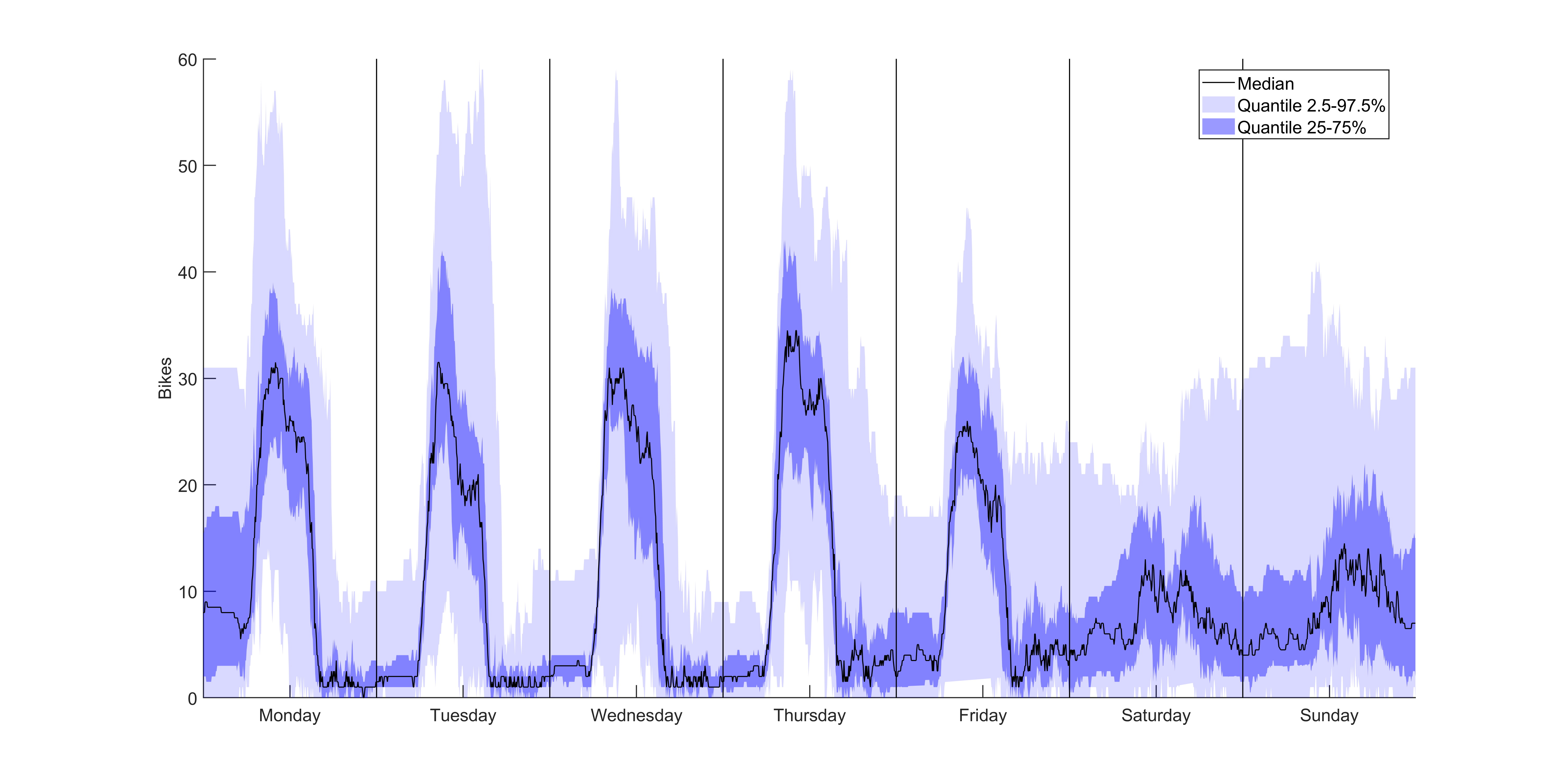}
  \caption{Station \textit{Itämerentori}.}
  \label{fig:fboxplot_ita}
\end{subfigure}
\begin{subfigure}{0.49\textwidth}
  \centering
  \includegraphics[width=1\linewidth,trim= 100 0 100 0]{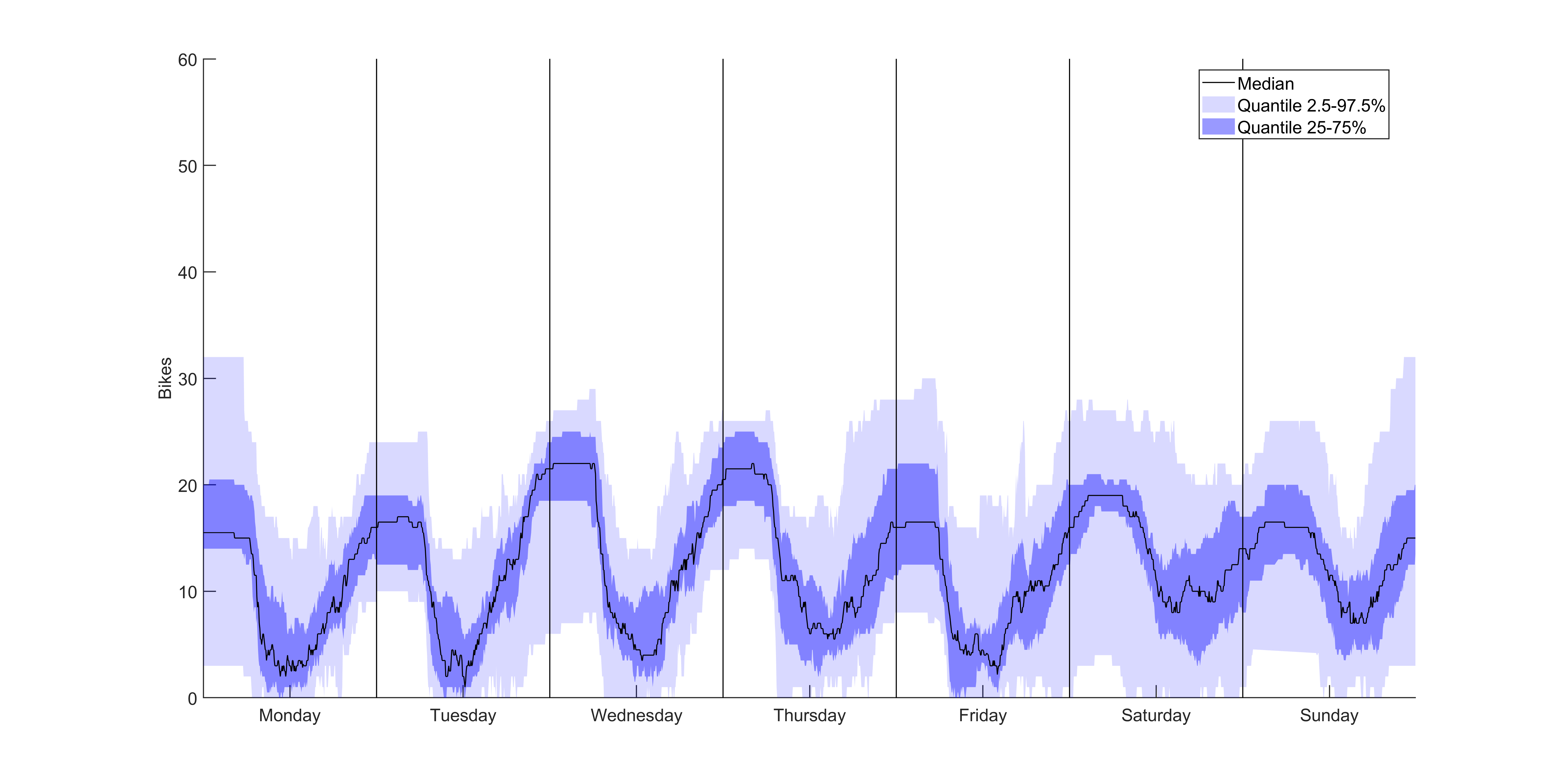}
  \caption{Station \textit{Haukilahdenkatu}.}
  \label{fig:fboxplot_hauki}
\end{subfigure}
\caption[Functional boxplots.]{Functional boxplots of two \bs\ stations summarising their spatiotemporal functional observations.}
\label{fig:fboxplot_both}
\end{figure}

To analyse the temporal dependence, functional boxplots are used (cf. \citealt{sun2011functional}). Figure~\ref{fig:fboxplot_both} show these plots for the stations \textit{Itämerentori} (Figure~\ref{fig:fboxplot_ita}) and \textit{Haukilahdenkatu} (Figure~\ref{fig:fboxplot_hauki}). Both functional boxplots highlight the periodic behaviour belonging to the cycle of one day. Both stations are characterised by a change in the number of bikes allocated during the morning hours from 7 to 10 and another major change in the afternoon hours from about 14 to 18\clock{}. However, the directions of change, as  well as the ranges of observed bikes, are different. At \textit{Itämerentori} station, the number of bikes increases in the morning and decreases in the afternoon. In addition, up to 58\, bikes were observed at maximum. The opposite happens at \textit{Haukilahdenkatu} station. There is a decrease in the number of bikes in the morning and an increase in the afternoon.  Here, the maximum number of bikes observed  was 32\,bikes.

\begin{figure}
\centering
\begin{subfigure}{0.32\textwidth}
\includegraphics[width=1.0\textwidth]{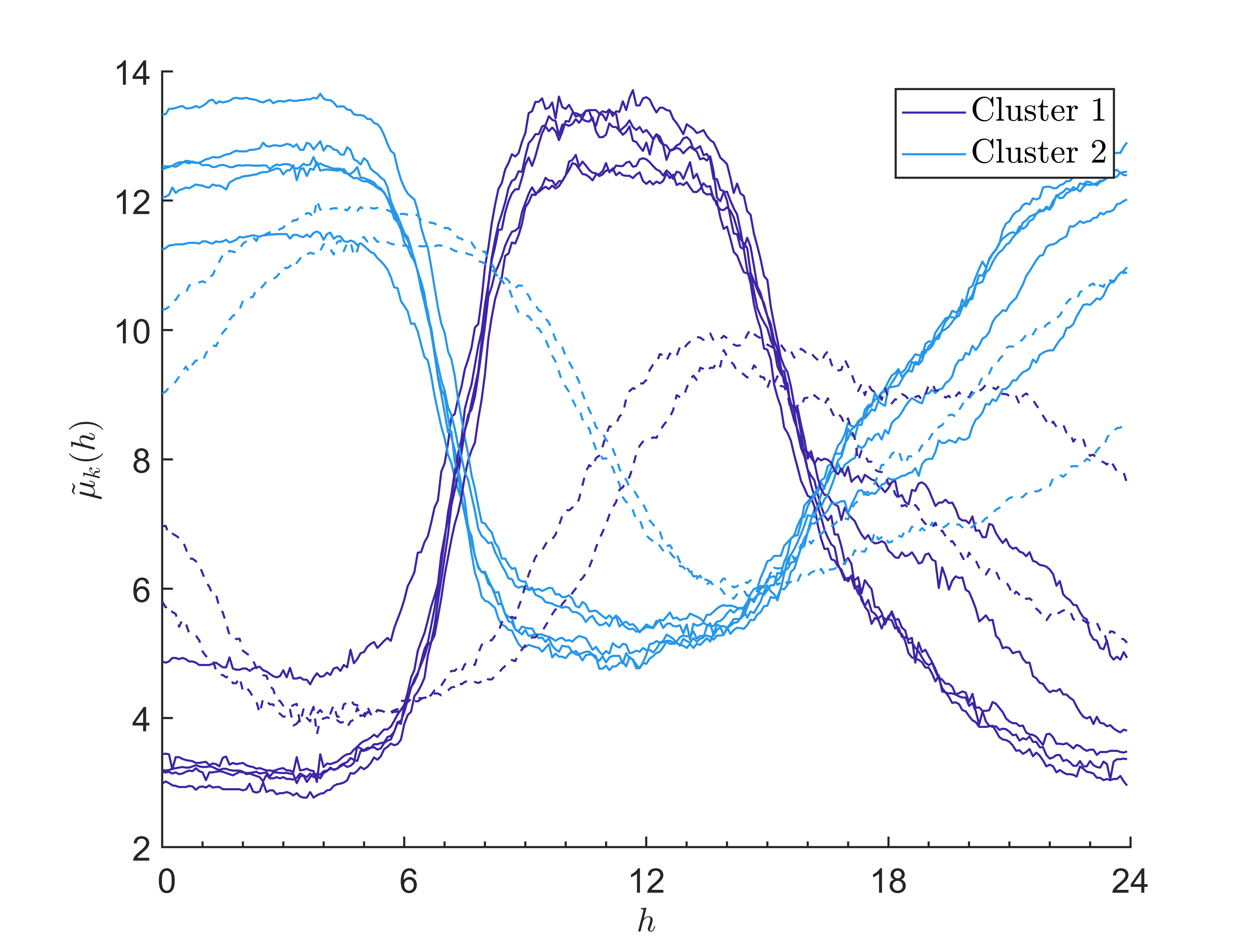}
\caption{Centres of the two clusters for each day of the week. Monday through Friday centres are shown with solid lines, while the Saturday and Sunday centres are depicted by dashed lines.}
\label{fig:kmeans_allclustercenters}
\end{subfigure}
\begin{subfigure}{0.32\textwidth}
\centering
\includegraphics[width=1\textwidth]{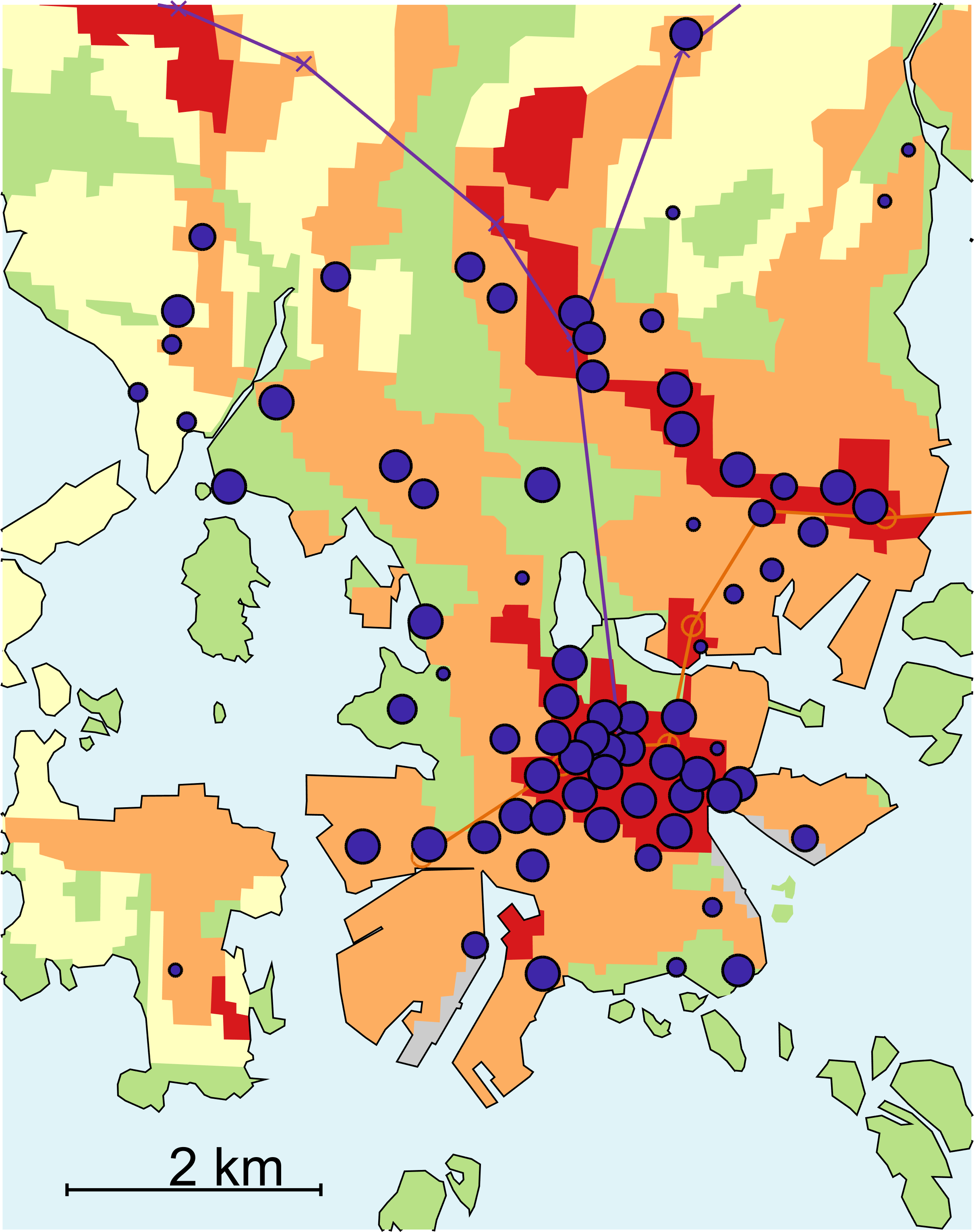}
\caption{Spatial allocation of Cluster 1.}
\label{fig:kmeans_final_cluster1}
\end{subfigure}
\begin{subfigure}{0.32\textwidth}
\centering
\includegraphics[width=1\textwidth]{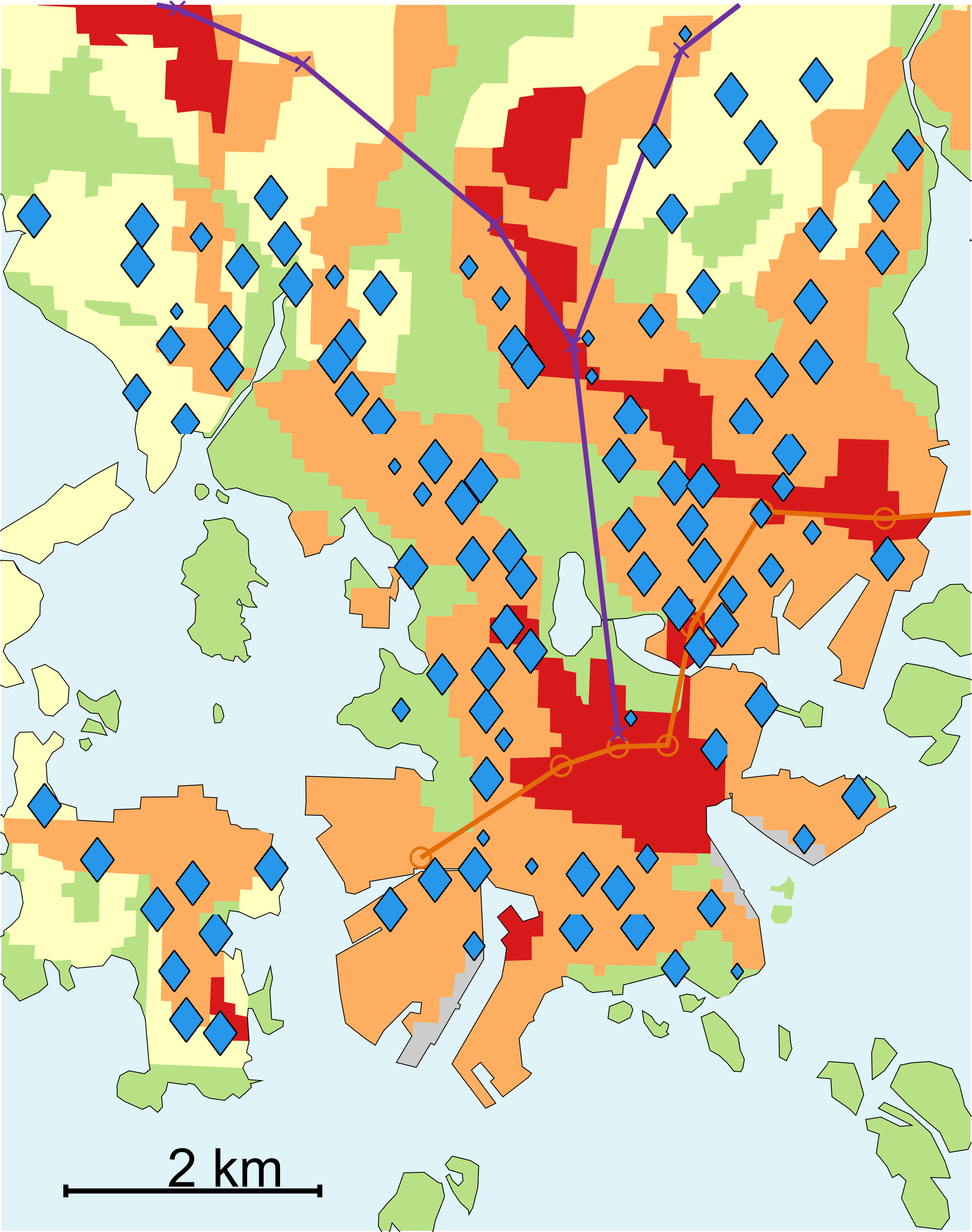}
\caption{Spatial allocation of Cluster 2.}
\label{fig:kmeans_final_cluster2}
\end{subfigure}
\caption{Assignment of the bike-sharing stations to the two clusters.}
\label{fig:kmeans_cluster1and2}
\end{figure}

Because we observed these two different type of stations, we performed a preliminary cluster analysis. To be precise, $k$-means clustering was applied to the median function of all observations, ensuring that these clusters are robust against outliers. Moreover, we chose Pearson's correlation as a  distance measure between the median curves and the cluster centres.
The clustering was conducted separately for each day of the week. Hence, the clustering yielded two cluster centres~$\tilde{\mu}_k\h$ for each day of the week. These centres are shown in Figure~\ref{fig:kmeans_allclustercenters}. The solid lines represent the cluster centres~$\tilde{\mu}_k\h$ for Monday through Friday, and the cluster centres for the weekend are denoted by dashed lines. The weekend’s dashed lines are similar to each other but differ from the solid lines for Monday through Friday by a positive shift on the time axis of approximately four hours. Additionally, the magnitude of change in the function is less for the cluster centres for the weekend than for the cluster centres~$\tilde{\mu}_k\h$ for Monday through Friday. Due to doing separate clusters for all days of the week, a station could be assigned to different clusters over the course of one week. However, for all days, the numbers of stations assigned to the clusters are similar, with roughly 40\,\% of the stations belonging to the first cluster and consequently about 60\,\% to the second cluster. Using these percentages, we can also ensure the correct cluster proportions for the bootstrap samples in the following analysis.

The assignments are shown in Figures~\ref{fig:kmeans_final_cluster2}~and~\ref{fig:kmeans_final_cluster2}, where the sizes of the symbols denoting the locations of the stations are scaled by the number of assignments out of seven to that respective cluster. The colours used for the areas correspond to the four city land use categories assigned by the 2016 Helsinki master plan for city development \citep{CityHelsinki2020}. These four categories consist of the city centre (orange), shops (red), recreational areas (green) and predominantly residential areas (beige). Taking a closer look at the locations of both clusters, one can see that type 1  stations (i.e., stations from cluster 1) are mostly located in the centre close to shopping areas, whereas the type 2 stations (i.e., stations from cluster 2) are outside of these shopping areas. That is, people use the bike-sharing system to commute to these centres during the day. Thus, many bicycles are available during day, while the opposite is true for the residential areas (i.e., the type 2 clusters).

\subsection{Results and interpretation}

In the following section, the focus is on the results of the empirical analysis. More precisely, we fitted a spatiotemporal functional model with different covariates and interaction effects. The model setup (i.e., the choice of the spatial covariance function, splines basis, knots and so on.) was determined via a cross-validation study using 1000~bootstrap iterations and a sample size of 30 stations for both the in-sample and out-of-sample cases. Note that the choice of size is a trade-off between reliability and computational complexity. The stations were drawn from the first and the second cluster according to the proportions described above (i.e., $41.4\,\%$ for cluster 1, and $58.6\,\%$ for cluster 2). Importantly, $\xvec{b}_{\text{in-sample}} \cap \xvec{b}_{\text{out-of-sample}}$ is the empty set.

In our case, an exponential covariance function fits the data the best. Hence, there are no anisotropic dependencies, which would indicate a prevalent direction of bike usage. Furthermore, for the basis function expansion, the B-splines approach was chosen, although the spectral time series analysis revealed periodic structures in the time series for all the stations. However, the B-splines allow the break points to be adapted according to the variation in the data along the functional dimension. Hence, the break points are selected to focus on the high variation during the daytime in such a manner that the standard deviation~$\tilde{\sigma}\h$ of the modelling error~$\varepsilon\sth$ remained less then 3\,bikes. To be precise, the position of the break points were set to
\begin{eqnarray}
\text{break points} = \{0, 5, 7.14, 9.29, 11.43, 13.57, 15.71, 17.86, 20, 24\}~\text{\clock} \notag
\end{eqnarray}
with only a few B-splines supporting the morning and evening hours, as there is little variation in the spatiotemporal functional observations during these periods. In contrast, there is high variation in the functional observations during the middle of the day; thus, the break points are denser during this period.

\begin{figure}
\centering
\includegraphics[width=0.8\textwidth]{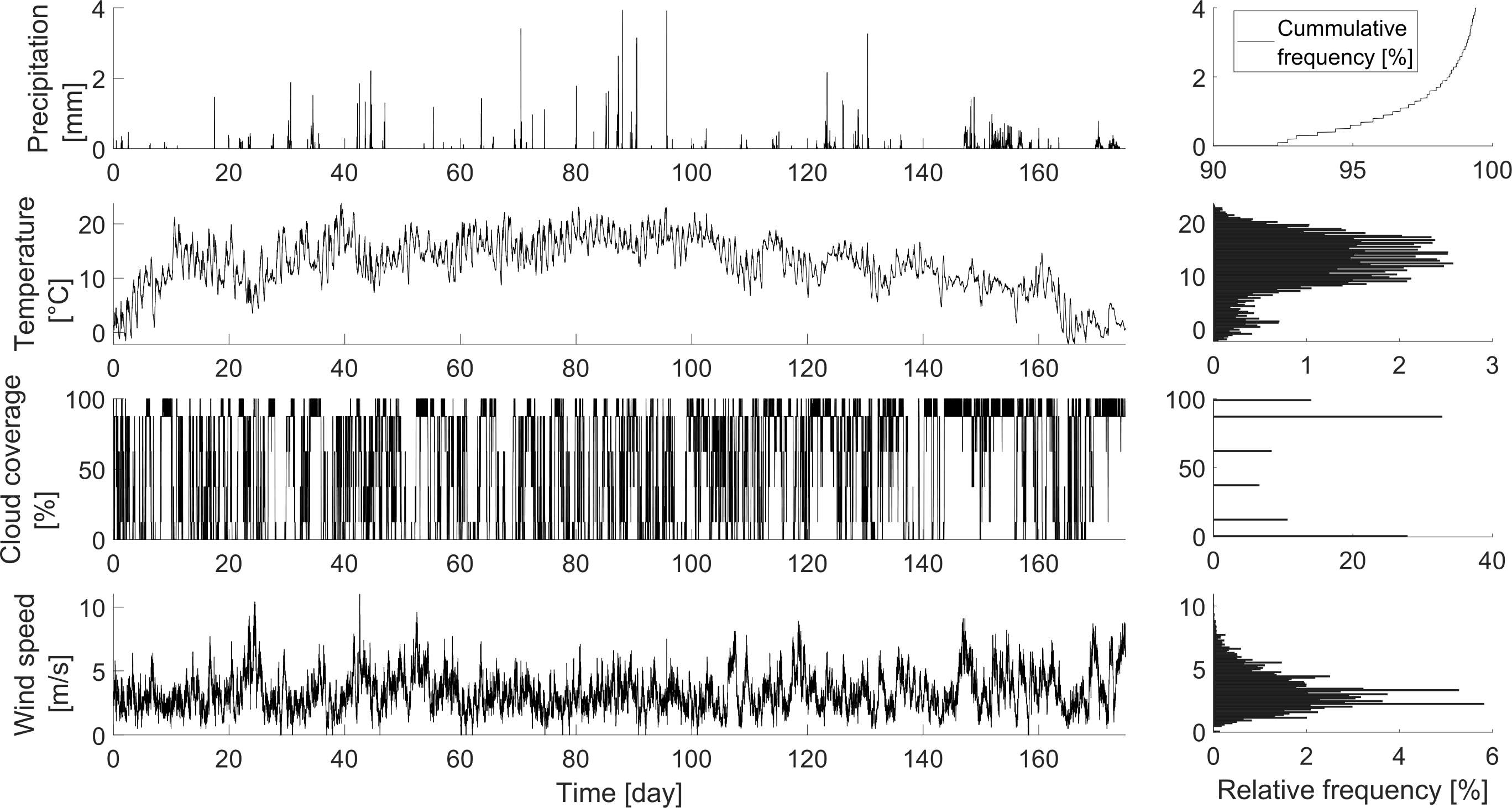}
\caption[Time series and histograms for four meteorological variables from 9th May until 30th October, 2017.]{Time series and histograms for four meteorological variables from 9th May until 30th October, 2017. The histograms on the right-hand side refer to the time series on the left-hand side. A cumulative histogram is shown for precipitation.}
\label{fig:weather}
\end{figure}

In addition, we have included several meteorological covariates. Three observatories are located within the spatial extent of the bike-sharing stations in Helsinki, but we only have used data from the Kaisaniemi observatory located to the northeast of the city centre. The spatial differences in these weather covariates are neglectable; hence, the observations are assumed to be constant over space but not time. Figure~\ref{fig:weather} shows the four weather covariates for the period covering the $9^\text{th}$~of~May to the $31^\text{st}$~of~October, 2017. Further, the respective histogram shows the distribution of the meteorological observations with the relative frequencies of occurrence. The histograms on the right-hand side of Figure~\ref{fig:weather} are aligned with the observations over time on the left-hand side.

To estimate different effects for each cluster, we also include interaction effects. That is, there are two intercept functions, one for each cluster, and all regressive effects are estimated separately. In the final model, we include dummy variables for Saturday and Sunday showing the weekend effects, meteorological variables (i.e., temperature, cloud coverage, wind speed, precipitation), a geographical variable, namely the elevation of the station, and infrastructure variables (i.e., distance to closest metro and train station). Doing so leads to two intercept functions and nine interactions for each cluster; in total, 20 parameter functions must be estimated.

\subsubsection{Fixed effects}

\begin{figure}
\centering
\includegraphics[width=0.49\textwidth]{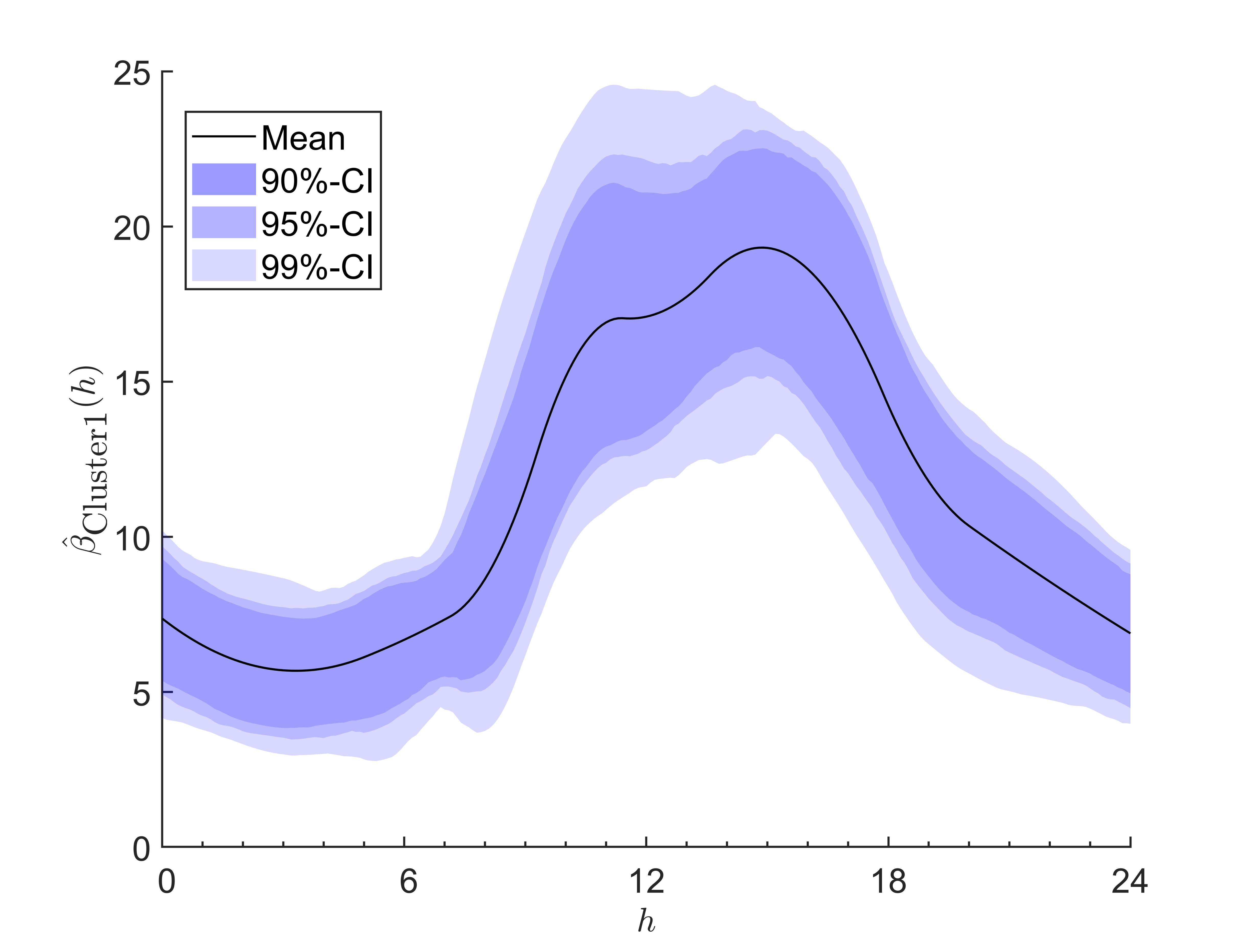}
\includegraphics[width=0.49\textwidth]{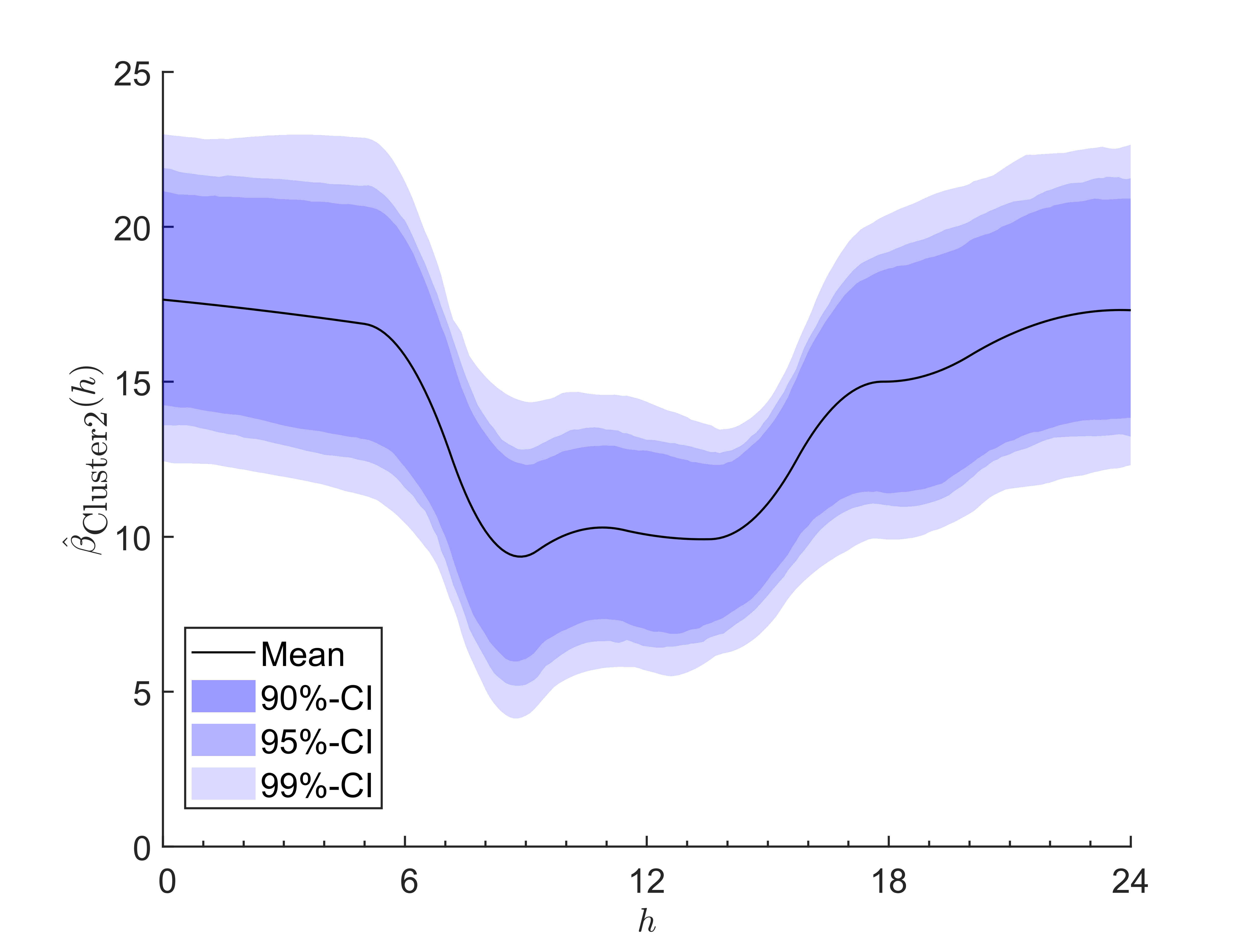}
\caption{Intercept: estimated functional intercepts for \bhat{Cluster1} (left) and \bhat{Cluster2} (right).}
\label{fig:intercepts}
\end{figure}

The two intercepts referring to the stations' cluster memberships are shown in Figure~\ref{fig:intercepts}. The mean curve of \bhat{Cluster1} shows about 7\,bikes during the night and in the evening, while the number of bikes increases during the day and has a first peak at approximately 11\clock{}, with about 17\,bikes, and a second peak at 15 in the afternoon, with 19\,bikes. The \ci\ has the widest range at the peaks. As expected, \bhat{Cluster2} shows the opposite shape. In the beginning and the end of the day, approximately 17\,bikes are located at stations from the second cluster. The number grows smaller during the day, with the minimum of 9\,bikes occurring at approximately 9 in the morning. Between 10 and 14\clock{}, there are 10\,bikes; afterwards, the number of bikes increases again.
For both intercepts, the \cip{95} along the entire function indicates a range of possible values of up to $\pm 2$\,bikes around the mean.

\begin{figure}
\centering
\includegraphics[width=0.49\textwidth]{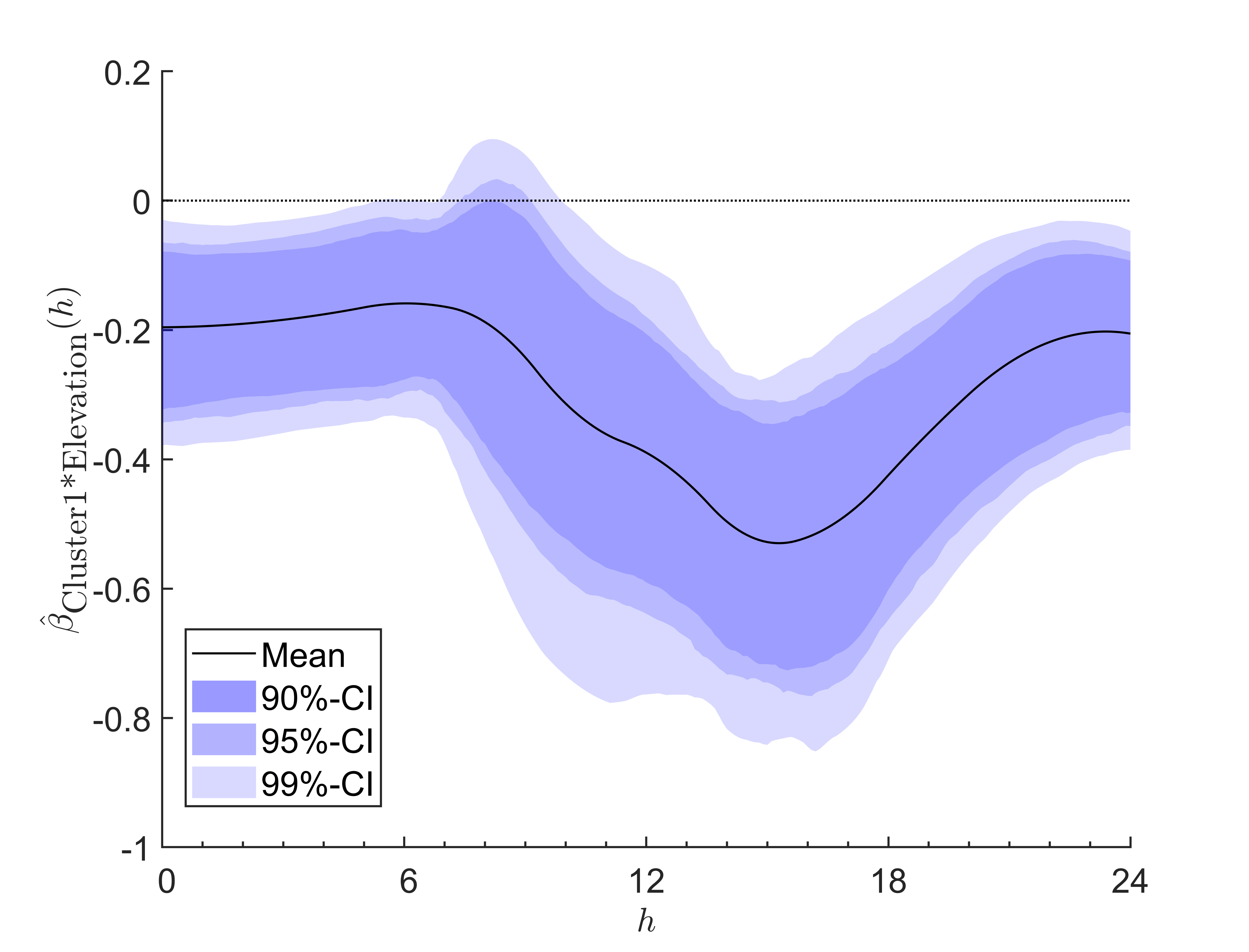}
\includegraphics[width=0.49\textwidth]{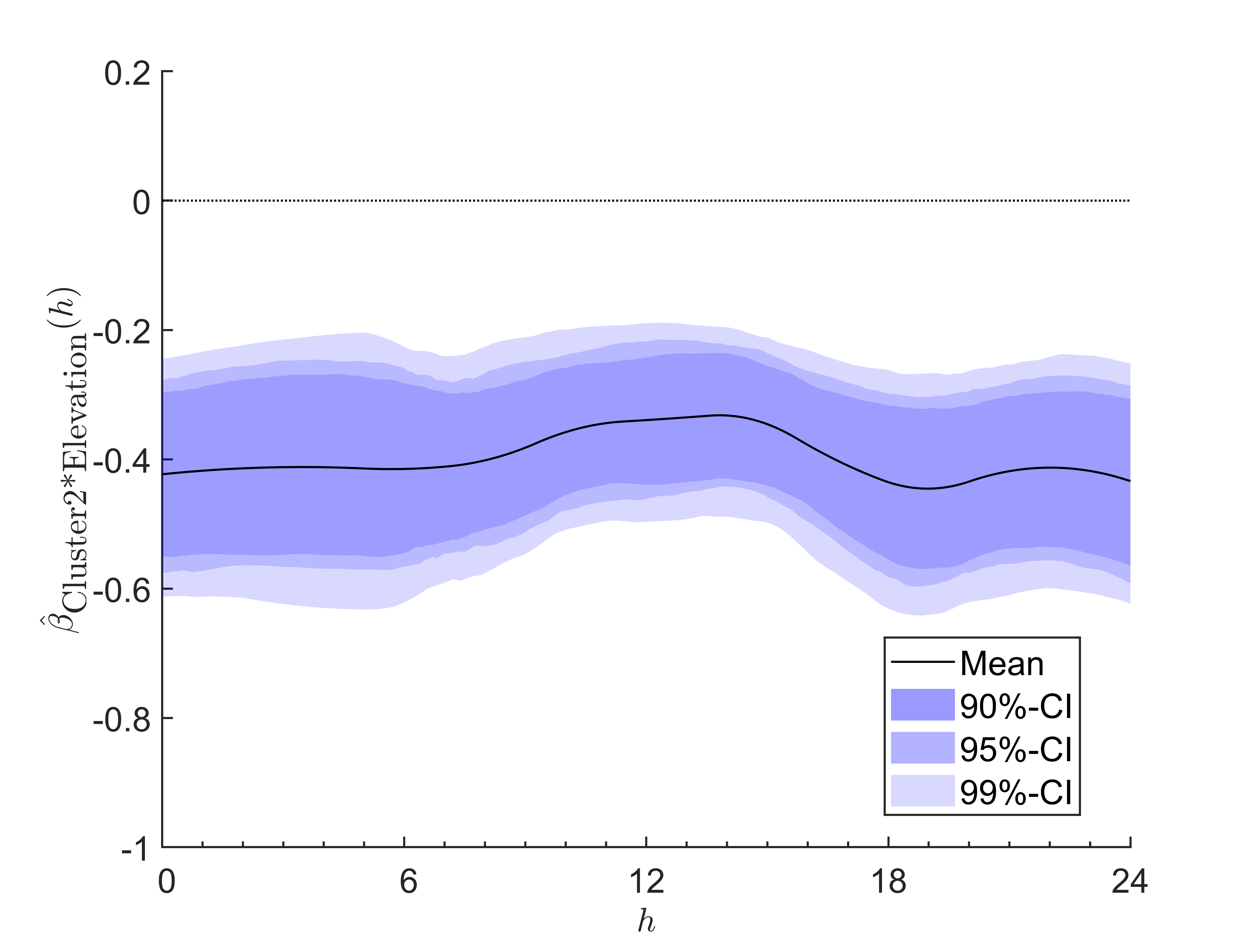}
\caption{Elevation: estimated functional influences of \bhat{Cluster1*Elevation} (left) and \bhat{Cluster2*Elevation} (right).}
\label{fig:full2_elevation}
\end{figure}

The influence of a \bs\ station's elevation on the number of allocated bikes is given by~\bhat{Cluster1*Elevation} and~\bhat{Cluster2*Elevation}, as shown in Figure~\ref{fig:full2_elevation}.
Both functions are significant and have negative signs along their entire domains. The influence of the interaction of elevation with the second cluster is around $-0.4\,{\text{bikes}}/{\text{m}}$. There is little variation about the mean, which does not change significantly. In comparison, the interaction \bhat{Cluster1*Elevation} varies more. Due to the negative sign, the number of allocated bikes at a station decreases as the elevation of the station increases, showing empirically that cyclists prefer cycling downhill over cycling uphill. The change in \bhat{Cluster1*Elevation} in the afternoon emphasises that cyclists might use the bikes even less for cycling uphill in their free time.

\begin{figure}
\centering
\includegraphics[width=0.49\textwidth]{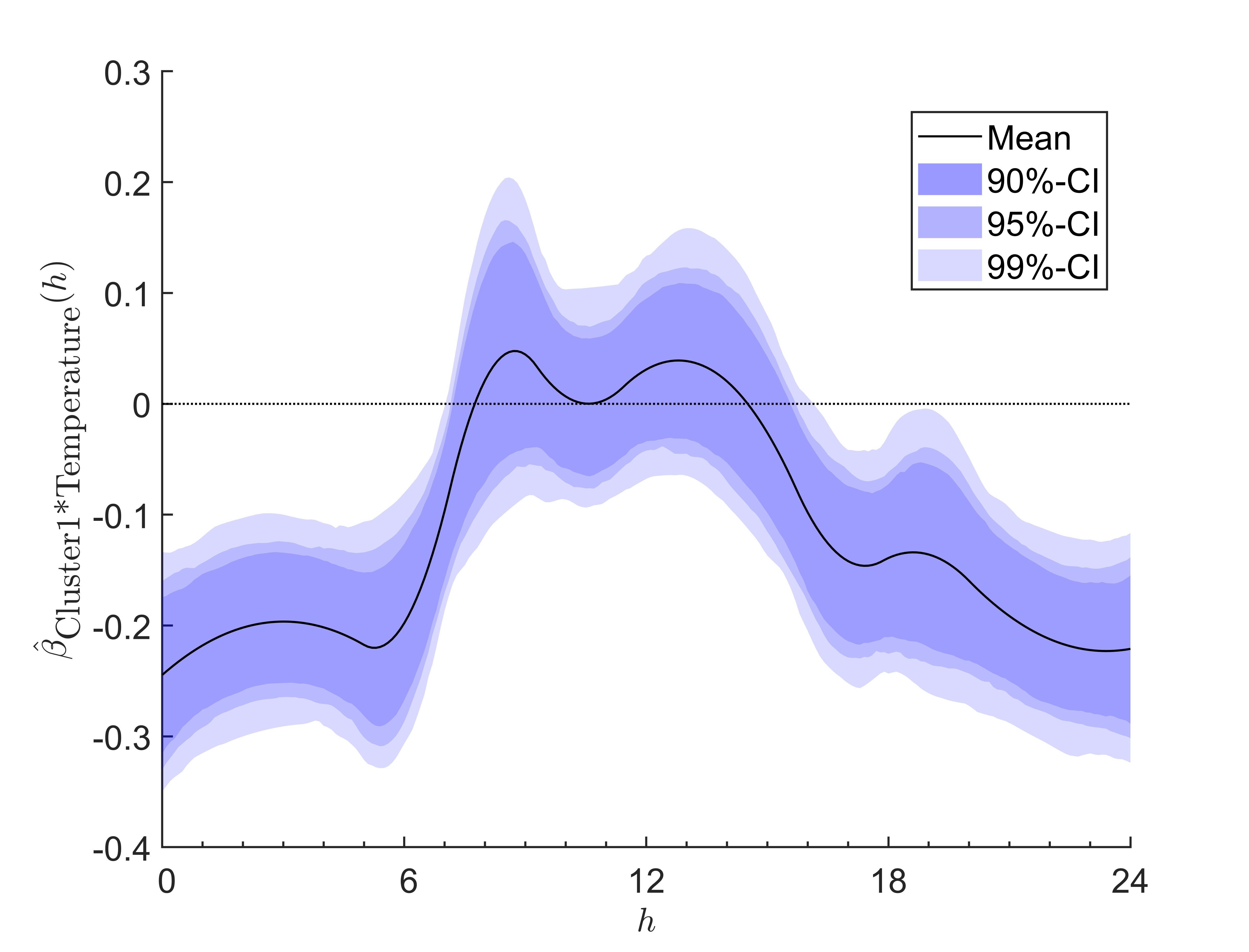}
\includegraphics[width=0.49\textwidth]{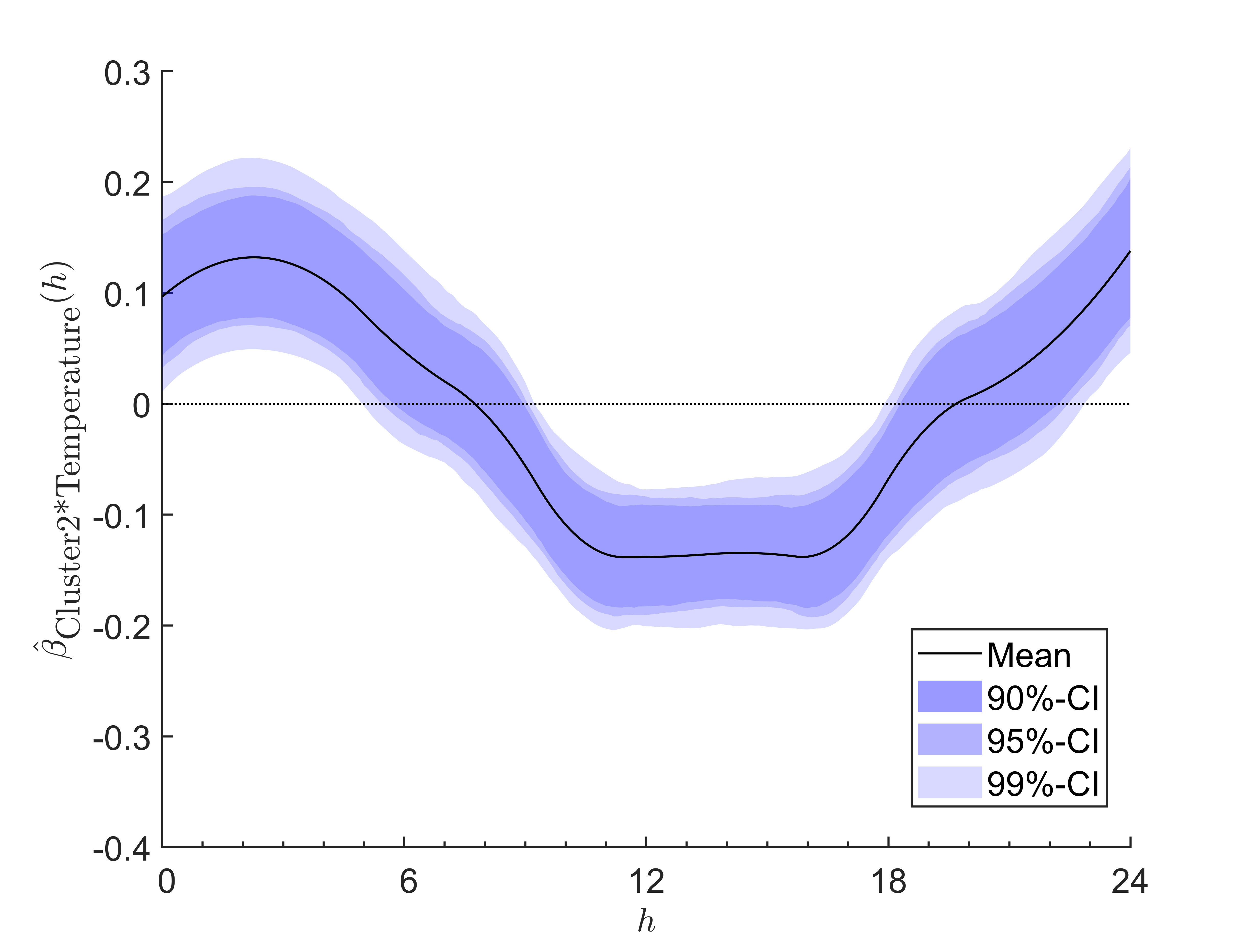}
\caption{Temperature: estimated functional influences of  \bhat{Cluster1*Temperature} (left) and \bhat{Cluster2*Temperature} (right).}
\label{fig:full2_temperature}
\end{figure}

In Figure~\ref{fig:full2_temperature}, the interactions with the intercepts of the first cluster \bhat{Cluster1*Temperature} and the second cluster \bhat{Cluster2*Temperature} are depicted. Regarding cluster 1, we observe that the number of allocated bikes changes by $-0.2\,{\text{bikes}}/{^\circ \text{C}}$ in the night and evening. Between 8 and 15\clock{}, the influence of the temperature is not significantly different from zero. As a consequence, the higher the temperature in Helsinki, the fewer bikes are located at the stations in the first cluster in the evening and night, while a temperature change between 8 and 15\clock{} has no significant effect on the number of bikes. For cluster 2, the interaction starts in the night at around $0.1\,\frac{\text{bikes}}{^\circ \text{C}}$, decreases rapidly to $-0.15\,{\text{bikes}}/{^\circ \text{C}}$ from 11 to 16\clock{} and increases afterwards up to $0.14\,{\text{bikes}}/{^\circ \text{C}}$. The shape of the function is similar to the valley-like shape of the intercept \bhat{Cluster2}. Consequently, the higher the temperature in Helsinki, the more bikes are located at the stations in the second cluster in the night and evening, while the number of allocated bikes decreases during the daytime.

Both interactions of the temperature with the intercepts yield functions with shapes similar to that of the function of the intercept itself, meaning that an increase in the temperature amplifies the already existing mountain-and-valley-like shape of the intercepts. The effect of the temperature shows that usage of the bike-sharing scheme is higher when it is warmer, as the change in the number of allocated bikes at stations from both clusters increases.

Most surprising are the results regarding the effect of precipitation as no effects could be identified from the data. However, bear in mind that only 8\,\% of all precipitation observations were non-zero. Including precipitation as a dummy variable could perhaps enable to see an effect for precipitation. Furthermore, public transport was included in the model, as \cite{jappinen2013modelling} suggested that it was one of the major factors influencing \bs\ usage. The influence of the distance of a \bs\ station to public transport varies. While the estimate~\bhat{Cluster2*Train} is not significantly different from zero, \bhat{Cluster1*Train} shows significant effects from 6 to 8 in the morning and from 15 to 17 in the afternoon. In the morning, the number of bikes increases by about $3\,{\text{bikes}}/{\text{km}}$ with increasing distance from the closest train station. The opposite is the case in the afternoon, with $-2.2\,{\text{bikes}}/{\text{km}}$ with increasing distance from the closest train station.
The intercept of the first cluster increases during the respective morning period and decreases in the afternoon. This change is amplified by the interaction~\bhat{Cluster1*Train}, meaning that more bikes are allocated to \bs\ stations further away from the train stations. This finding supports the hypothesis that commuters use the public bike-share scheme to overcome distances from the train station to their destinations, also known as the last-mile problem . Moreover, the greater the distance of a \bs\ station from the closest metro station, the more bikes are allocated to this station during the morning until the mid-afternoon. Thus, stations closer to the public transportation system are used more frequently (i.e., there are fewer bikes allocated during the day). Interestingly, the influence of the distance to metro stations is only less than half the effect of the distance to train stations.

\begin{figure}
\centering
\includegraphics[width=0.49\textwidth]{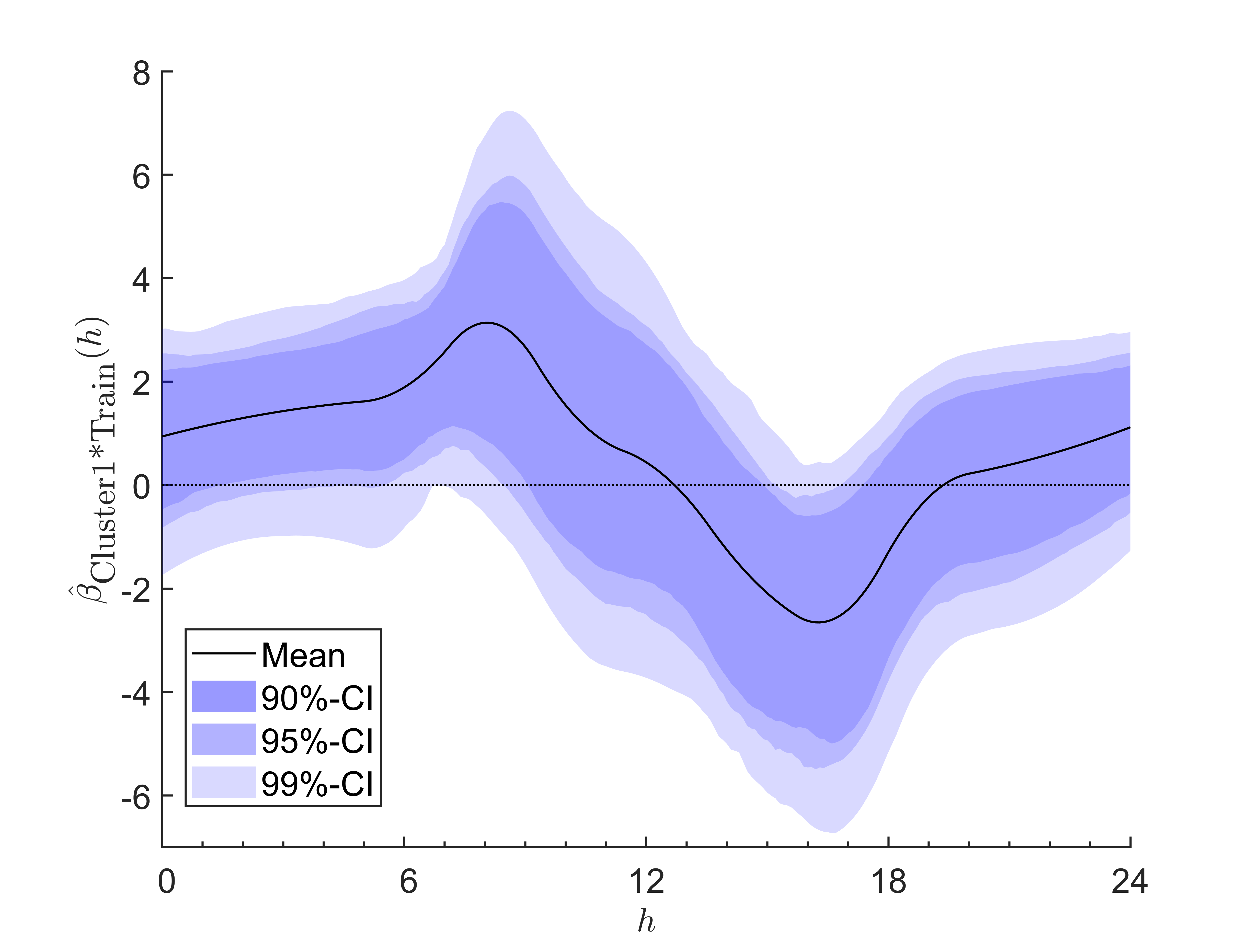}
\caption{Distance to train station: estimated functional influence of  \bhat{Cluster1*Train}.}
\label{fig:full2_cluster1_train}
\end{figure}

\subsubsection{Random effects}

\begin{figure}
\centering
\includegraphics[width=0.49\textwidth]{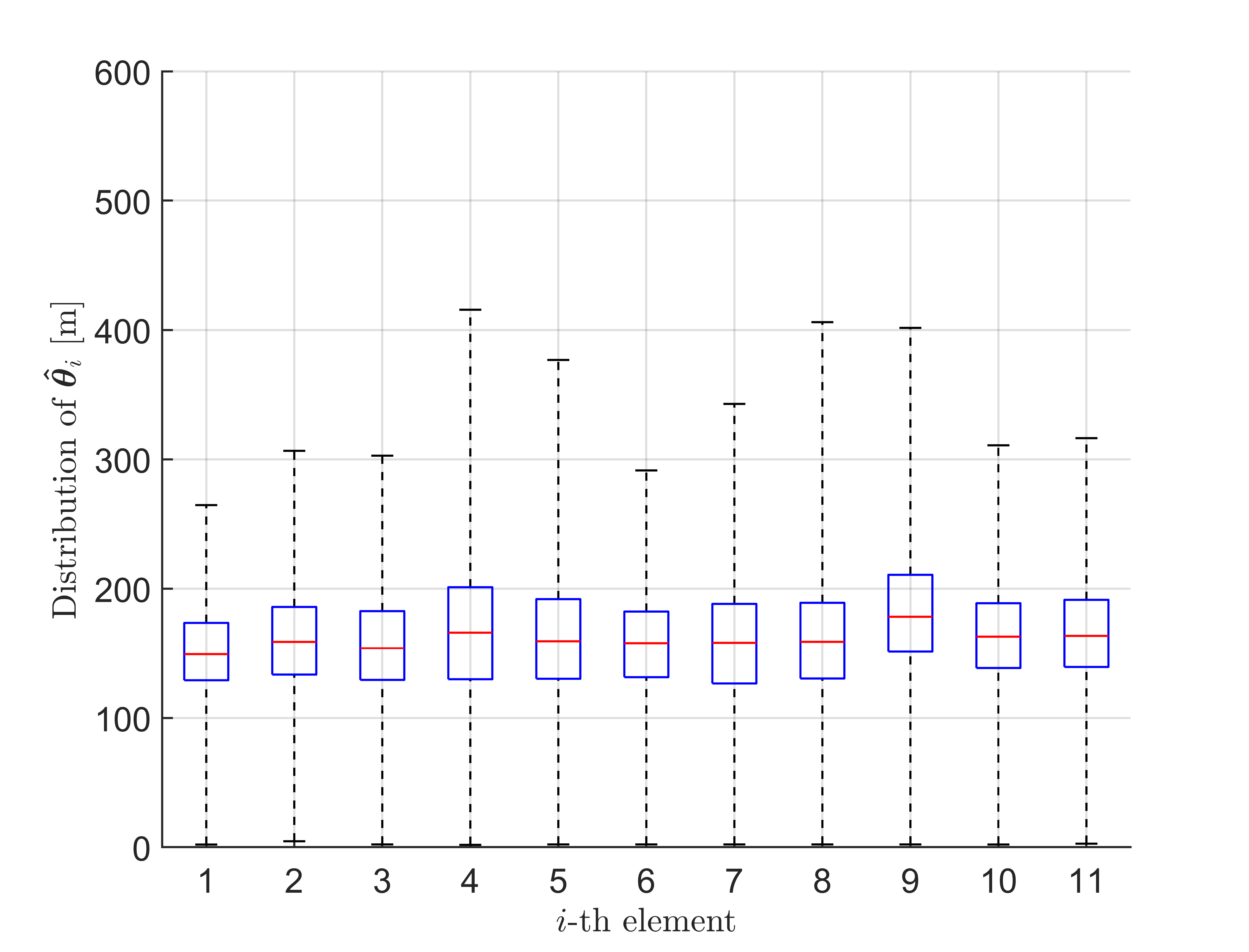}
\caption{Distribution of the estimated range parameters~$\hat{\xvec{\theta}}$. The boxplots show the median in red, the 50\,\%-central region is shown in blue and the whiskers denote the minimum and maximum values of the distribution.}
\label{fig:full2_theta}
\end{figure}

Below, we discuss the results of random effects and the error term briefly. The random-effects model explains the random variation in the data that is not explained by the fixed effects, i.e., the mean of the process and the covariates. First, the temporal autoregressive dependence is estimated with the diagonal elements of the transition matrix~$\hat{\xmat{G}}$. The median values of the bootstrap estimates of these diagonal elements range from roughly 0.45 to 0.50, where all elements are similar. The minimal and maximal values are in the ranges of $[0.27,0.38]$ and $[0.54,0.64]$, respectively. Consequently, about half of the random variation observed in a day is explained by the previous day. To analyse the spatial dependence, Figure~\ref{fig:full2_theta} shows the estimated range parameters~$\hat{\xvec{\theta}}$. Their median values are around 160\,m. Since the exponential covariance function declines rapidly and the covariance is about~0.37 at $h=\theta$, the spatial dependence of the process is weak in most cases - the largest values are in the range $[264,415]$\,m. Considering the distances between the bike-sharing stations in Helsinki, where the median distance is 3\,km, the estimated range parameters $\hat{\xvec{\theta}}$ reveal that the spatial dependence is constrained stations that are very close together. Whereas, the elements of the cross-correlation matrix~$\hat{\xmat{V}}$ indicate a smaller random effect at night than during the day, it seems that the degree of the spatial dependence is mostly stable, with slightly more variability occurring during the day.

\subsubsection{Out-of-sample forecasts}

\begin{figure}
\centering
\includegraphics[width=0.49\textwidth]{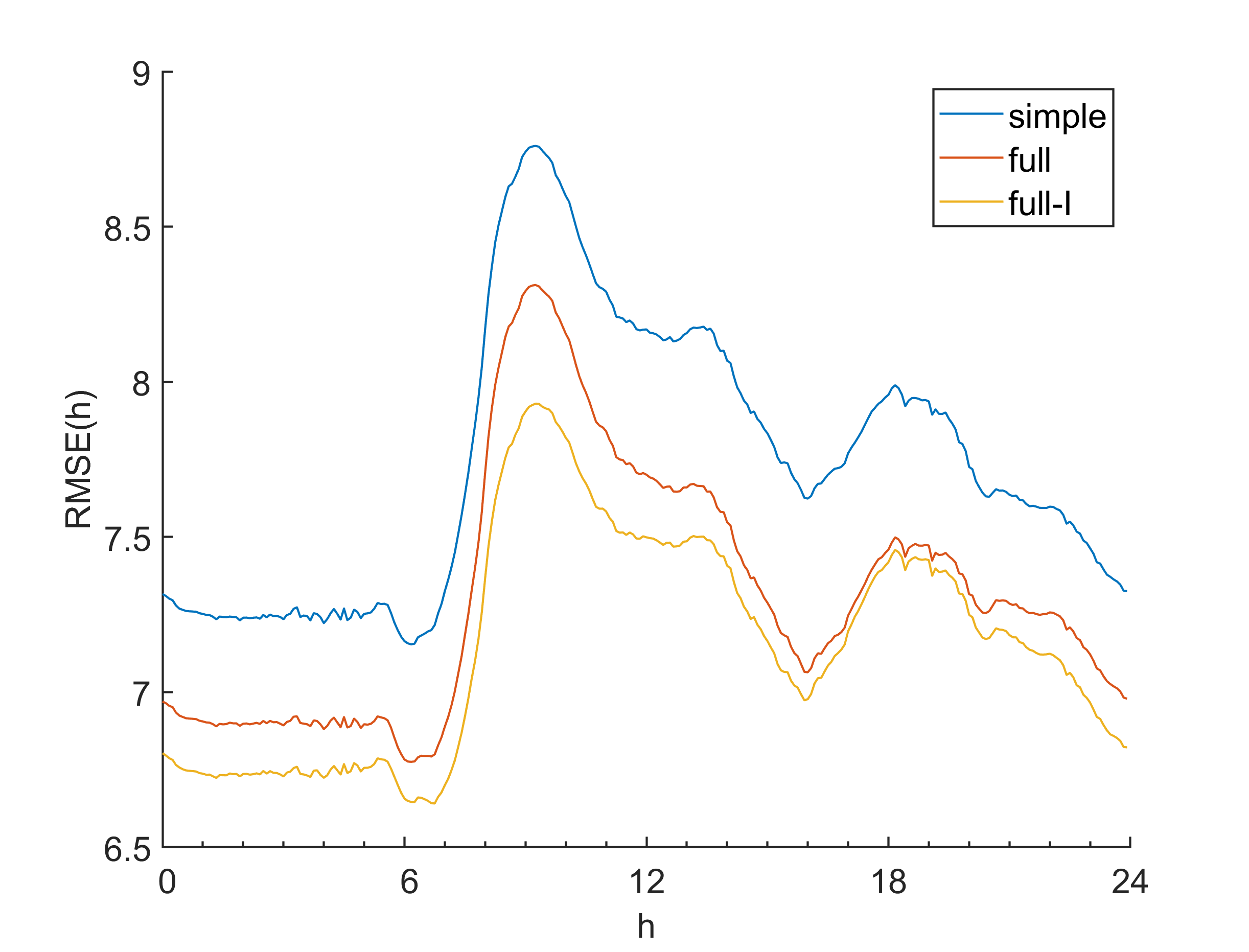}
\includegraphics[width=0.49\textwidth]{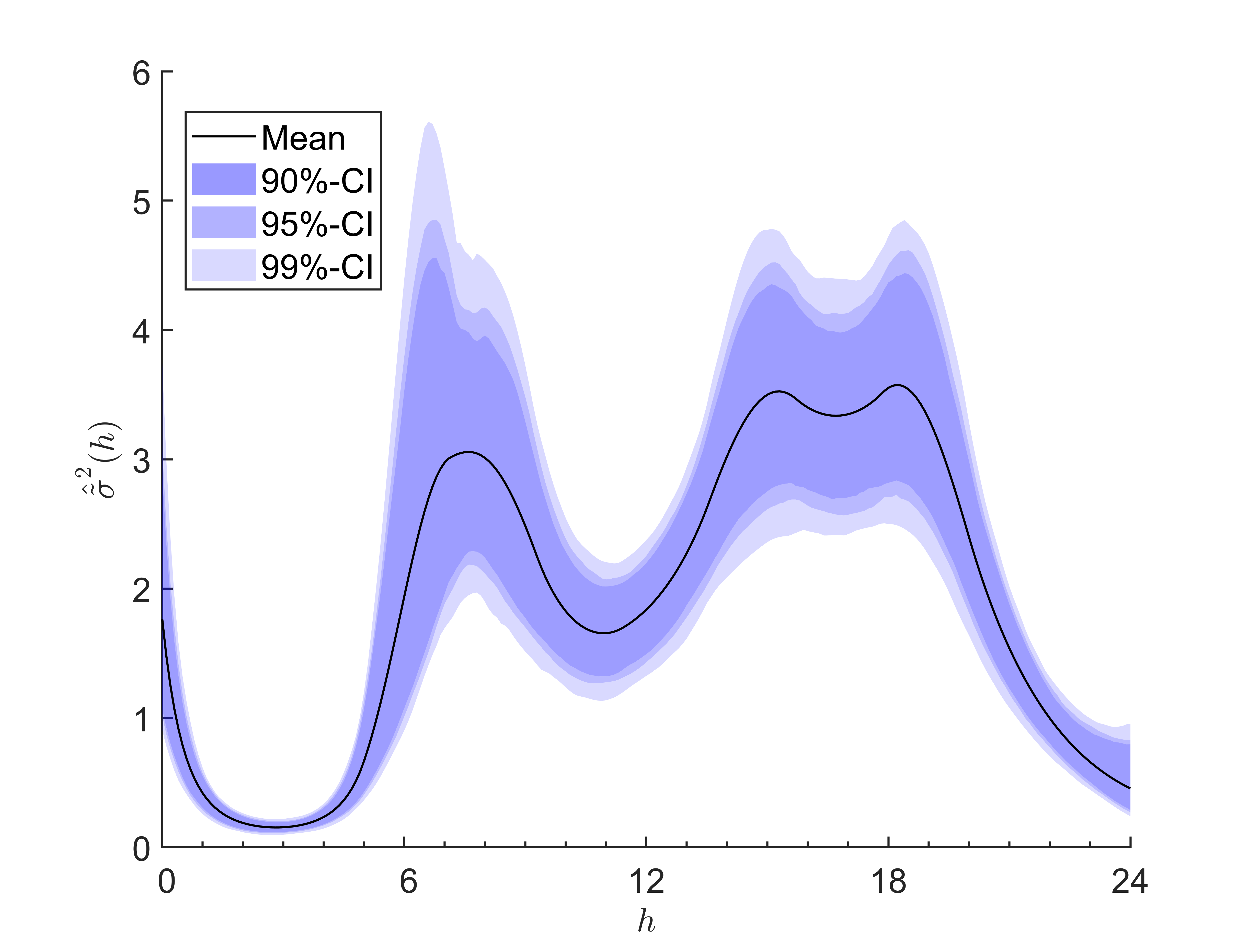}
\caption{Comparison of the functional $\text{RMSE}(h)$ (left) values for three selected models (blue: intercept-only, red: full model without interactions, orange: full model with interactions) and the estimated functional measurement variances $\hat{\tilde{\sigma}}^2(h)$ (right).}
\label{fig:rmse_h_comparison}
\end{figure}

Finally, one question remains open - how well does the model fit the data? The out-of-sample RMSE is depicted in Figure~\ref{fig:rmse_h_comparison} with an orange line. For comparison, two alternative models (i.e., a simple intercept-only model, and a model with all regressors but no interactions with the clusters) are shown with blue and red curves. Interestingly, the three functions have roughly the same shape but are shifted. The RMSE is the lowest from midnight to 7\clock{} and increases drastically between 7 and 9\clock{}. After reaching a maximum at 9 in the morning, it decreases until 16\clock{} and has a local maximum between 17 and 19\clock{}. The range is $\pm 1.5$\,bikes over the course of the day. Considering the daily out-of-sample RMSE, we observe median values varying between 5 to 9 bikes. Moreover, the estimated functional error variance $\hat{\tilde{\sigma}}^2(h)$ is shown in Figure~\ref{fig:rmse_h_comparison}. Across the entire day, the maximal standard error is less than 2\,bikes. During the night from about 1\clock{} to 5\clock{}, the variance is the smallest, showing that the variation in the data for all \bs\ stations is low in the night. Also, from 22 in the evening until midnight, the standard deviation is less than 1\,bike. In contrast, the variance is higher in the morning from 6 to 9 and in the afternoon from 15 to 19\clock{}. Here, the corresponding standard deviation is up to about 1.9\,bikes.

\section{Conclusion}\label{sec:conclusion}

This paper focused on station hire data from the \bss\ in Helsinki. With over 7 million observations from 140 bike-sharing stations taken at five-minute intervals, the analysis of bike-sharing  station usage was brought to a new level, as the entire complex dataset was considered and knowledge about the changes in the influencing factors over the course of a day was inferred. We simultaneously accounted for spatial, temporal and spatiotemporal dependence by applying a hierarchical statistical model in a functional framework. The model parameters were estimated using the implemented maximum-likelihood approach of the software package D-STEM (see \citealt{finazzi2014dstem}). To supply computationally efficient estimated standard errors and guarantee a certain robustness against outliers, a bootstrap approach was applied.

Most findings about the influencing factors were in line with the results from existing literature, although comparability is limited due to the use of a different methodology and data. We have shown that the bike-sharing stations can be divided into two clusters depending on the similarities in their spatiotemporal functional observations. It is important to note that these similar functional observations cluster together in space too. The estimated parameters have shown that the morning rush hour is particularly difficult to model and predict. There is a mountain-like shape to the daily available bikes for stations belonging to predominantly working areas. By contrast, we observe a valley-like shape in living areas. This behaviour is different on weekends, where the daily peaks are also shifted towards the afternoon. Furthermore, we examined which weather conditions could have an influence. According to \cite{eren2020review}, precipitation should affect \bs\  station usage the most among the weather conditions. Here, however, the influence of precipitation was not significant.

A drawback of the model can be seen in the out-of-sample RMSE values , which ranged between 7 and 9 bikes, giving it a range similar to that for the random effects. The random effects covered the unexplained variation in the station hire data. Unfortunately, out-of-sample validation of models from the literature were not found, meaning that the error mentioned above could not be compared and evaluated.
To better understand the implications of different types of the bike-sharing station usage, future studies could address spatially-varying covariates, perhaps providing insights into the cause of the separation into clusters. On the other hand, \bss\ station usage is governed by the decisions made by individuals and maybe even pure coincidences in their behaviour. In general, analysing the relationship between the number of allocated bikes at the \bs\ stations and the proposed covariates produces a correlation and does not necessarily imply causality.

It is worthwhile to develop and apply complex models to spatiotemporal functional data from \bsss, as detailed knowledge can be gained and perhaps lead to future improvements in implemented \bsss.


\end{document}